\documentclass[aps,titlepage,12pt]{revtex4}
\usepackage{amsfonts}
\usepackage{amsmath}
\usepackage{graphicx}
\usepackage{dcolumn}
\usepackage{bm}
\usepackage{overpic}
\usepackage{booktabs}
\usepackage{color}
\usepackage{tabularx}
\usepackage{multirow}

\begin{document}
\title{Disruptive papers in science are losing impact}

\author{An Zeng$^1$\footnote{anzeng@bnu.edu.cn}, Ying Fan$^1$, Zengru Di$^1$, Yougui Wang$^1$, Shlomo Havlin$^{2,}\footnote{havlin@ophir.ph.biu.ac.il}$}

\affiliation{
$^1$ School of Systems Science, Beijing Normal University, Beijing, China.\\
$^2$ Department of Physics, Bar-Ilan University, Ramat-Gan 52900, Israel.
}

\begin{abstract}
The impact and originality are two critical dimensions for evaluating scientific publications, measured by citation and disruption metrics respectively. Despite the extensive effort made to understand the statistical properties and evolution of each of these metrics, the relations between the two remain unclear. In this paper, we study the evolution during last 70 years of the correlation between scientific papers' citation and disruption, finding surprisingly a decreasing trend from positive to negative correlations over the years. Consequently, during the years, there are fewer and fewer disruptive works among the highly cited papers. These results suggest that highly disruptive studies nowadays attract less attention from the scientific community. The analysis on papers' references supports this trend, showing that papers citing older references, less popular references and diverse references become to have less citations. Possible explanations for the less attention phenomenon could be due to the increasing information overload in science, and citations become more and more prominent for impact. This is supported by the evidence that research fields with more papers have a more negative correlation between citation and disruption. Finally, we show the generality of our findings by analyzing and comparing six disciplines.
\end{abstract}

\date{\today}

\maketitle

\section{Introduction}
Innovation is a key feature for scientific research~\cite{the2017zeng,science2018fortunato}. In order to accelerate innovation in science, much effort has been made to evaluate originality of scientific studies and understand the behavioral factors related to scientific creativity~\cite{age2011jones,collaboration2005uzzi,incentives2011azoulay}. In recent years, the disruption index has been proposed and widely used to quantify the originality of individual study~\cite{dynamic2017funk}. With the help of this metric, it has been found that smaller~\cite{large2019wu}, fresher~\cite{fresh2022zeng} and flatter~\cite{flat2022xu} teams are usually associated with more original studies. It has also been revealed that disruption is usually smaller in larger fields~\cite{slowed2021chu} and scientific papers are generally becoming less disruptive over time~\cite{papers2023park}.

Another important feature for evaluating scientific research is the impact of the resultant papers, which has been investigated for much longer time~\cite{review2016waltman}. The most straightforward way to quantify a paper's impact is to use its citation count~\cite{an1998redner}. Other metrics, such as impact factors for journals~\cite{ciation1972garfield} and H-index for researchers~\cite{an2005hirsch} are all based on paper citations. Despite its simplicity and wide usage, there are consistent efforts in overcoming the drawback of using the citation count~\cite{review2016waltman}, resulting in a variety of new metrics such as rescaled citations for cross-disciplinary comparison~\cite{universality2008radicchi} and $C_{10}$ for studying time evolution~\cite{quantifying2016sinatra}. There exist also the PageRank algorithm~\cite{finding2007chen} and its variants~\cite{pagerank2015gleich,google2016ermann} aiming at quantifying the impact of papers by taking into account their neighbors' impact, realized by iterative processes in the citation networks. In recent years, increasing attention has been paid to understand the mechanism of impact formation~\cite{quantify2013wang,impactful2022zeng}. Examples include the mechanistic model for characterizing the citation evolution of individual papers, as well as the random-impact rule~\cite{quantifying2016sinatra} and hot-streak phenomenon~\cite{hot2018liu} for the most influential works in scientists' careers.

Despite the existing efforts in understanding the impact and innovation of scientific publications, the relation between these two aspects of impact and disruption remains unclear. So far, there are very few related works in the literature. As innovation is to some degree connected to combination of distant knowledge~\cite{invention2015youn,interact2010sood}, the revealed connection between atypical combination and scientific impact suggests a possible positive correlation between innovation and impact~\cite{atypical2013uzzi}. However, an investigation on scientific teams shows that smaller teams are associated with higher disruption and lower impact~\cite{large2019wu}, suggests a possible negative correlation between innovation and impact. Some recent studies suggest that combining citations and disruption can result in a better identification of Nobel laureates~\cite{evaluate2023wang} and Nobel prize-winning papers~\cite{quantify2023wei}. In this context, some key questions naturally arise. How the innovation and impact of scientific papers are related? In other words, do highly innovative papers tend to receive more attention? With the accelerating growth of scientific papers, does the relation between innovation and impact change over time?

To address these issues, we systematically study the evolution of the correlation between disruption and citation of individual papers, finding that these two metrics become more negatively correlated over the years. Furthermore, we find that citation of disruptive works becomes lower and lower than average during the years, and disruption of highly cited papers decreases faster than disruption of less cited papers. We also use reference age, reference popularity and reference diversity as alternative measures of paper originality. We find that papers citing older references, less popular references and diverse references yield less citations. The increasing of negative correlation between citation and disruption over the years is found to be probably related to the increasing information overload in science. Indeed, we find that research fields producing more papers have a stronger negative correlation between citation and disruption. We finally show that our finding of consistent increase of negative correlations is general and appear in six different disciplines, including physics, computer science, chemistry, biology, social science and multidisciplinary science.

\section{Results}
We begin by introducing the two key metrics considered in this paper, i.e. citation and disruption. The ways to compute these two metrics are illustrated in Fig. 1. Citations of a focal paper are simply the number of papers citing it. Papers published earlier have longer time to accumulate citations. Thus, to make papers published at different time comparable, throughout this paper we use $C_5$ to measure the impact of a paper, namely the number of citations received in 5 years since its publication. The disruption is a measure of the originality of a paper. The value of disruption of a paper is from $-1$ to $1$, see detailed definition in Methods. A larger disruption of a paper reflects that more papers cite it and do not cite its references, corresponding to higher originality. To be consistent with the $C_5$ index, we only consider the citing papers in 5 years since the publication of the focal paper when computing its disruption, and denote it as $D_5$. Fig. 1 illustrates also the possible relations between these two metrics. The light orange regions in the x-y plane denote the possibility of positive correlation between citation and disruption, meaning that highly original papers tend to have high impact. The light blue regions denote the possibility of negative correlation between citation and disruption, meaning that highly original papers tend to have little impact. The first question we asked is what is the real correlation between citation and disruption.

To answer this question, we analyze the scientific publication data of 6 distinct research fields, including Physics, Computer Science, Chemistry, Biology, Social Science and Multi-disciplinary Science. The data set of each field consists of the scientific papers published in the major journals representing the field. A full description of these data sets is provided in the Methods. Throughout the paper, we will mainly present the results in the field of Physics based on the data of the American Physical Society (APS) journals, containing 482,566 papers ranging from 1893 to 2010. The results for the other fields are similar to those of Physics, and are summarized in Fig. 6 and Supplementary Figs. 9-13.

We first present in Fig. 2a the scatter plot of the citation $C_5$ versus disruption $D_5$ values of papers, with a curve marking the average. One can see that the dots spread all over the plane and the curve is rather flat, both suggesting that these two metrics over all years are almost uncorrelated (with Pearson correlation coefficient $-0.05$). In Fig. 2b, we analyze the papers published in a given year and compute the (Pearson) correlation between these papers' citation and disruption, and denote it as yearly correlation. After obtaining the yearly correlation for every year, we show the distribution of the yearly correlations in Fig. 2b. In addition, we show the result of a randomized surrogate where the publication years of papers are randomly reshuffled. The results show that despite the narrow distribution of the randomized surrogate, the empirical correlation in some years can be significantly outside the surrogate distribution e.g., even larger than $0.1$ or smaller than $-0.1$. While studying the average year of the data points in the tails of the distribution (i.e. $\leq-0.1$ and $\geq0.1$), we find that the large correlations in the right tail happen around 1958 while the small correlations in the left tail happen around 2003. This prompts us to investigate how the correlation between citation and disruption evolves with years. Indeed in Fig. 2c, we calculate the average disruption $D_5$ of papers for increasing values of $C_5$. We compare the curves for the papers published before 1960 and the papers published after 2000. One can see interestingly that while these two metrics are positively correlated in the early years, they are negatively correlated in recent years. To confirm this difference, we directly show in Fig. 2d the evolution of yearly correlations over all studied years. We consider the evolution of correlations starting from 1950 since before 1950 there exists too little data to obtain robust statistics. It is seen in Fig. 2d that all three types of correlation coefficients (i.e. Pearson, Spearman, Kendall) exhibit clear decreasing trends. This result suggests that highly original papers tend to have less impact and receive less attention over the years.

We notice that papers with few citations may have extreme values of disruption simply because of the low number of citations, as shown in Fig. 2a. To avoid this artificial effect on the general trend, we calculate in Supplementary Fig. 2 the evolution of the correlation between disruption and citation only for papers with at least 10 citations. The decreasing trend of the correlations are still significant. To further support the observed trend, we also examine in Supplementary Fig. 2 the randomized surrogate case where the publication years of papers are randomly reshuffled. In the controlled surrogate, all the correlation coefficients become flat, which suggests that the original decreasing trend in empirical data is indeed a true pattern. Also, as the disruption value strongly depends on the number of references~\cite{papers2023park}, in Supplementary Figs. 3 we examine the results with the number of papers' references controlled. We obtained similar decreasing trends when the effect of these factors removed.

To better understand the decreasing correlation between citation and disruption during the years, we directly study the citations of disruptive papers. We first compare in Fig. 2e, the distributions of the bootstrap citation $C_5$ of all the papers before 1960 and the papers before 1960 with positive disruption $D_5>0$, respectively. One can clearly see that the disruptive papers before 1960 have higher citations than average. On the other hand, when we compare the distributions of the bootstrap citation $C_5$ for the papers after 2000, we find that the disruptive papers received less citations than average. To quantify this effect, we compute a metric called relative citation $C_5$ which is simply the mean citations $C_5$ of the papers published in a year with $D_5>0$ or $D_5<0$ divided by the mean citations of all the papers published in this year. We show the evolution of the relative citation in Fig. 2f. It is seen that the relative citation of $D_5<0$ papers steadily increases over the years while the relative citation of $D_5>0$ papers strongly decreases. Taken together, the disruptive papers are indeed gradually losing impact.

A recent paper has revealed that the disruption of papers decreases over the years~\cite{papers2023park}. We show in Fig. 3a the evolution of the disruption of all the papers, impactful (10\% most cited in $C_5$) papers, and less impactful (10\% least cited in $C_5$) papers in each year, respectively. One can see that the overall decreasing trend is consistent with ref.~\cite{papers2023park}, however, the disruption of impactful papers decreases much faster. The disruption of less impactful papers, on the other hand, exhibits a much less significant decreasing trend. Consequently, the disruption of impactful papers is higher than less impactful papers in the early years, while the opposite occurs in recent years. For statistical results, we analyze papers with different percentiles of $C_5$ in each year and compute in Fig. 3b the difference of the mean disruption of papers after 2000 and the mean disruption of papers before 1960. One can see that the decreasing trend indeed becomes less significant for papers at low $C_5$ percentiles. To support these findings, we show in Fig. 3c the evolution of the fraction of $D_5>0$ papers. Consistent with the trend in Fig. 3a, the fraction of $D_5>0$ papers decreases with years and this fraction among highly cited papers decreases much faster compared to other papers. The statistical results are shown in Fig. 3d where the decreasing fraction of $D_5>0$ papers indeed becomes less significant for papers with low $C_5$ percentiles. As the number of references of a paper is an important factor affecting its disruption value~\cite{papers2023park}, we study in Supplementary Fig. 4 the evolution of the disruption of impactful and less impactful papers, taking into account only the papers with similar number of references in the APS data. It is seen that the disruption of impactful papers still decrease over the years, while the disruption of less impactful papers shows even a slight increasing trend over time.

As the calculation of the disruption of papers involves the information of citing papers, one may wonder whether the decreasing trend of the correlation between disruption and citation is because these two metrics share the information of citing papers. In order to address this issue, we consider three other metrics that have been shown to connect to paper originality~\cite{large2019wu,atypical2013uzzi,papers2023park} but their calculation solely depends on paper references (thus completely independent of citations). These metrics are reference age, reference popularity and reference diversity. In general, a highly innovative paper tend to dig deeper in existing papers and thus may discover ideas or methods in older references and less popular references~\cite{large2019wu}. Furthermore, a highly innovative paper may tend to combine the knowledge from distant fields, and thus may cite diverse references~\cite{atypical2013uzzi,papers2023park}. The definitions of reference age and reference popularity are straightforward. The reference age is simply the mean difference between the publication year of the focal paper and its references. The reference popularity is the average $C_5$ of the focal papers' references. The reference diversity measures how atypical a papers' references are combined. One should first calculate a quantity as the reciprocal of one plus the frequency of each pair of a paper's references being co-cited in the literature, and the reference diversity of the paper is simply the average of this quantity over all reference pairs of the paper (for a mathematical definition, see Methods). The basic statistics of these metrics are given in Supplementary Fig. 5. In Fig. 4a-c, we show respectively the evolution over years of the correlation between papers' citation $C_5$ and its reference age, between papers' citation $C_5$ and it reference popularity, and between papers' citation $C_5$ and its reference diversity. One can see that these results are consistent with our earlier findings. The correlation between citation and reference age decreases (Fig. 4a), the correlation between citation and reference popularity increases (Fig. 4b), and the correlation between citation and reference diversity decreases (Fig. 4c) over the years. To further support these trends, we show in Fig. 4d the distributions of bootstrap citation $C_5$ of papers published before 1960 and half of these papers with higher reference age. For comparison, we show also the same distributions for papers published after 2000. Comparing the results in the early years and recent years, one can see that the papers with larger reference age (analogous to high innovation) tend to have lower and lower citation than average. In Fig. 4e and 4f, we show respectively similar results for papers with smaller reference popularity and papers with higher reference diversity, and observe similar decreasing citations than average over the years. These trends support again that highly innovative papers receive less and less citations.

The decreasing attention to disruptive works, represented by lower citations than average, is a complex phenomenon that might be caused by multiple factors. One possible factor is the information overload in science resulted from the increasing number of scientific papers. When facing a large number of papers published each year, scientists cannot read every paper to identify the highly disruptive works. Instead, citation is a simple index one can obtain easily in many venues, so new citations are more likely to be given to already cited papers rather than to innovative ones~\cite{the2009newman}. To test this hypothesis and detect this signal empirically, we study in Fig. 5a the number of published papers in each year, identifying an exponential growth of the number of new papers. In Fig. 5b, we analyze all the papers published in year 2000, and show the citations these papers received in the first 5 years after publication (from 2001 to 2005) $\Delta k_{\rm first}$ versus the citations received in the second 5 years after publication (from 2006 to 2010) $\Delta k_{\rm second}$. The diagonal averaged curve confirms the preferential attachment mechanism~\cite{emergence1999barabasi}. In order to quantify the significance of the preferential attachment effect, we directly study the Pearson correlation between $\Delta k_{\rm first}$ and $\Delta k_{\rm second}$, and show in Fig. 5c the evolution of the correlation by analyzing papers published in different years. The results show that as time evolves, the correlation keeps increasing. The inset shows that the yearly citation share of the top-1\% most cited papers increases over time~\cite{handful2012barabasi}. In Supplementary Fig. 6, we show also the yearly citation share of the top-10\% most disruptive papers decreases over time. It seems plausible that as the citation preferential attachment becomes stronger, the role of other factors (e.g. disruption) becomes less and less significant in attracting future citations.

The above results inspire us to examine research fields of different sizes within Physics. Research fields are naturally defined in the APS data by the PACS (Physics and Astronomy Classification Scheme) codes which are selected by authors to identify fields of their papers. The size of each field is simply the number of papers using the corresponding code. We show in Fig. 5d that the distribution of the field sizes exhibits an exponential form. The inset shows that the citation share of top-1\% most cited papers in a field is generally larger in bigger fields. In Fig. 5e, we investigate the correlation between citation $C_5$ and disruption $D_5$ of papers in fields of different sizes, finding that the correlation decreases with field size. In Fig. 5f, we show the relative citations $C_5$ for the 50\% papers with larger $D_5$ and for the 50\% papers with smaller $D_5$ in fields of different sizes. One can see that the more disruptive works in larger fields tend to have smaller number of citations than average. These results indicate that the higher overload of papers in larger fields indeed decreases the attention to disruptive works.

To test the generality of our finding of less attention to high innovation papers, we investigate data sets from six disciplines, including Physics, Computer Science, Chemistry, Biology, Social Science and Multidisciplinary Science. The results of these fields are summarized in Fig. 6 and Supplementary Figs. 9-13. In Fig. 6, we compare the Pearson correlation between citation $C_5$ and disruption $D_5$ in early years and in recent years. One can see a clear decrease in recent years of the correlation in all disciplines. The significance of the correlations can be clearly seen via a shifted correlation in Supplementary Fig. 7. In addition, we show the results of Spearman and Kendall rank correlation in Supplementary Fig. 8, both of which exhibit clear decreasing trends in all disciplines. In the lower panel of Fig. 6, we show also the relative citation of papers with positive disruption $D_5>0$ in the early years and in recent years. Consistent with the results in the upper panel of Fig. 6, one can observe a clear decrease of relative citations of disruptive papers over time. Taken together, the phenomenon of disruptive works losing attention seems to be universal across disciplines.

\section{Discussion}
In summary, we studied the relationship between innovation and impact of scientific publications, and find a significant decrease of their correlation to stronger negative values over the years. The citations of disruptive works becomes fewer than average in recent years, and the fraction of disruptive works keeps decreasing among highly cited papers. To further support this trend, we examine the evolution of the correlation by using innovation-related metrics that are solely based on paper references. We find a consistent decreasing trend of their correlation to the papers' impact. We also find indications that the more negative correlation between citation and disruption over the years is partially due to the fast increasing number of papers during the years, resulting in a stronger effect of citation preferential attachment. We finally compare six disciplines, including physics, computer science, chemistry, biology, social science, multidisciplinary science. In all studied fields we observe a consistent trend of losing impact over time in disruptive papers.

A very recent paper by Park et al.~\cite{papers2023park} finds surprisingly that scientific papers and patents are becoming less disruptive over time. Our finding here may have revealed the possible origin of this unexpected phenomenon. The information overload in science is getting increasingly serious nowadays. As a result, it becomes impossible for scientists to read all articles in their fields. The citation metric and its variants, which can be obtained easily and immediately, are usually used to select papers for detailed reading and referencing. Furthermore, citations play an increasingly significant role in science, motivated by the benefits in a variety of activities such as faculty hiring, funding application, and recognition (e.g. prizes). Therefore, many scientists try their best to increase the citations of their papers. At the same time, scientific journals aims to accept papers that potentially will have higher citations to increase their impact factors. Therefore, it is plausible that citation and disruption nowadays become negatively correlated, meaning that if one publishes disruptive work, its number of citations will be lower. In this context, the motivation to conduct highly original and innovative works gradually decreases, and results in the observed decrease of disruption of scientific papers over the years~\cite{papers2023park}.

There are several possible extensions that can be made based on this work. A straightforward extension is to investigate whether journals of higher impact factors indeed publish more innovative ground-breaking works. This may change the dominance of using impact factor for evaluating scientific journals and even papers~\cite{the2015hicks}. Also, the negative correlation between citation and disruptive also challenges the current system of using mainly citations to evaluate scientists. Therefore, metrics such as H-index~\cite{an2005hirsch} and total citations measure only the impact of scientists, not their creativity and innovation ability. Thus, a more comprehensive evaluation that naturally combines these two aspect is required. Our work is also of importance to decision makers. A possible conclusion of our study is that in order to encourage and motivate scientists to conduct more disruptive works, it might be more effective to promote the impact of disruptive works. Once the disruptive works could be promoted and obtain credit and support, they will probably receive higher attention. Such combined measure will encourage and motivate scientists to conduct disruptive works, and therefore the boundaries of scientific fields will be more effectively pushed forward.

\section{Methods}
\textbf{Data.} We study in this paper six large-scale data sets, including disciplines of physics, chemistry, biology, computer science, social science and multidisciplinary science. The physics data set consists of the scientific publications data of the American Physical Society
(APS) journals~\cite{quantifying2016sinatra}, with 482,566 papers ranging from year 1893 to year 2010. The computer science data is obtained by extracting scientists' profiles from online Web databases~\cite{extraction2008tang}. It contains 2,092,356 papers ranging from year 1948 to year 2014. The chemistry data contains the publications data of the American Chemical Society (ACS) journals, with 1,320,333 papers ranging from year 1879 to 2020. The biology data contains the publication data of Cell publishing group journals, with 154,233 papers ranging from year 2003 to year 2020. The social science data contains the publication data of SAGE publishing group journals, with 1,354,511 papers ranging from year 1965 to year 2020. The multidisciplinary science data contains all papers in five representative multidisciplinary journals including Nature, Science, Proceedings of the National Academy of Sciences (PNAS), Nature Communications and Science Advances. The dataset consists of 633,808 papers ranging from year 1869 to year 2020. The data of chemistry, biology, social science and multidisciplinary science are extracted according to the DOI of papers from a large publication data set freely downloaded from Microsoft Academic Graph~\cite{an2015sinha}.\\

\textbf{Disruption.} Disruption is an index measuring the originality of individual papers or patterns proposed~\cite{dynamic2017funk}. It is a local metric using only the information of the neighboring nodes of a focal paper in the citation networks to evaluate its originality. The basic idea is that a highly disruptive paper should have less of its citing papers that cite its references, and a consolidating paper otherwise, see Fig. 1. To calculate the disruption of a focal paper, one should first calculate the difference between the number of its citing papers that do not cite its references and the number of its citing papers that cite its references. The disruption index is obtained by dividing this difference by the number of its citing papers plus the number of subsequent papers of the focal paper that do not cite it but do cite its references. Accordingly, disruption varies between -1 and 1. A larger disruption indicates a higher originality of the paper. Note that the papers with no citation or no reference are excluded from our analysis, because their disruption cannot be evaluated. As the value of a paper's disruption depends on its citing papers which usually growth with time, we only take into account the citing papers and subsequent papers within 5 years of its publication to ensure focal papers published at different years comparable. We denote the index as $D_5$. The results of $D_{10}$ and $D_{15}$ are similar to those of $D_5$, and are presented in Supplementary Fig. 1.\\

\textbf{Reference diversity.} It has been pointed out that a highly innovative paper is associated with atypical combination of existing knowledge~\cite{invention2015youn,interact2010sood}, that is, bridging references that are usually not cited together~\cite{atypical2013uzzi}. Based on this concept, we consider an index called reference diversity. Denoting the historical co-cited times of a focal paper's two references ($i$ and $j$) as $n_{ij}$, the reference diversity of this focal paper is simply defined as the average of $1/(1+n_{ij})$ over each pair of its references. The reference diversity index ranges from 0 to 1, with a higher value indicating a paper's references are less frequently co-cited in the existing literature.\\

\textbf{Bootstrap citations.} We compare the mean citations $C_5$ of highly disruptive works published in a year and the mean citations $C_5$ of all papers published in that year. The bootstrap process can illustrate whether the difference between the means are sufficiently significant. The bootstrap citations was obtained by random sampling of papers' citations such that each paper's citation has an equal chance to
be selected and can be selected over and over again. The distributions were obtained by performing 10,000 realizations of bootstrap citations. A complete non-overlapped distribution supports the significant difference of the two means.\\

\noindent \textbf{Data availability.} The data sets used in this paper are publicly available. The APS data are available upon request submitted to \url{https://journals.aps.org/datasets}, the AMiner data can be freely downloaded via \url{https://www.aminer.cn/aminernetwork}, and the Microsoft Academic Graph data can be accessed in Zenodo via \url{https://doi.org/10.5281/zenodo.2628216}.\\

\noindent \textbf{Code availability.} Computational codes for data processing and analysis are available from the corresponding authors on request.

\clearpage
\noindent  \textbf{Acknowledgments.}\\
This work is supported by the National Natural Science Foundation of China under Grant(72274020 and L2224029). This work was also supported by the European Union under the Horizon Europe grant OMINO (grant number 101086321). Views and opinions expressed are however those of the author(s) only and do not necessarily reflect those of the European Union or the European Research Executive Agency. Neither the European Union nor European Research Executive Agency can be held responsible for them.\\

\noindent \\ \textbf{Author contributions.} \\
AZ, SH designed the research, AZ performed the experiments, YF, ZD, YW contributed analytic tools, AZ, SH analyzed the data, all authors wrote the manuscript.\\

\noindent \textbf{Competing interests.}\\
 The authors declare no competing interests.\\

\clearpage
\section{Figures}
\vspace{-0.5cm}
\begin{figure}[h!]
  \centering
  \includegraphics[width=16cm]{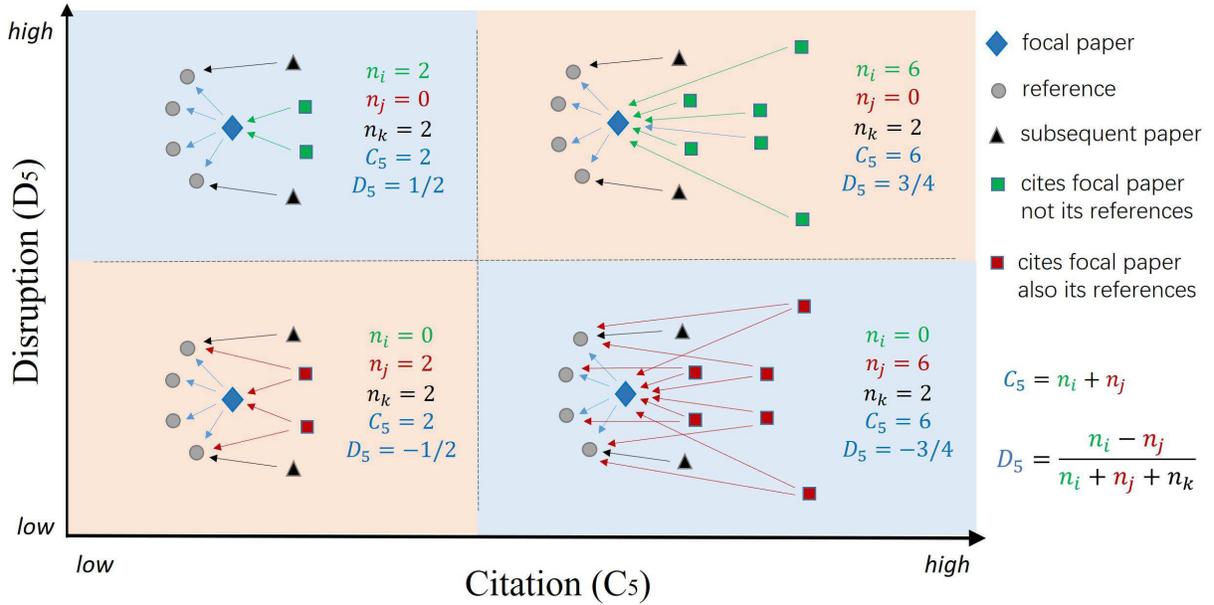}\\
  \caption{\textbf{Illustration of the possible relations between citation and disruption.} The x-axis shows the number of citations of a focal paper, while the y-axis shows the value of the disruption index of the paper. We divide the x-y plane into four regions, and in each region we use a toy citation network to demonstrate the computation of the corresponding citation and disruption values of a paper. Each toy network consists of a focal paper (diamond), its references (circles) and its citing papers (squares). The subsequent papers (marked as triangles) are the papers that do not cite the focal paper but cite its references. As the focal papers published earlier have longer time to accumulate citations, we only consider the citing papers and subsequent papers in 5 years since the publication of the focal paper. The citations $C_5$ of the focal paper are simply the number of papers that cite it, measuring the impact of the paper. The disruption $D_5$ of the focal paper is a ratio that measures the originality of the paper. To calculate the disruption of the focal paper, one should first calculate the difference between the number of its citing papers that do not cite its references and the number of its citing papers that cite its references. The disruption is obtained by dividing this difference by the number of all its citing papers plus the number of subsequent papers. The four regions have two background colors, where the light orange regions denote a possibility of positive correlation between citation and disruption (highly original papers tend to have high impact), and the light blue regions denote a possible negative correlation between citation and disruption (highly original papers tend to have little impact).}\label{fig1}
\end{figure}

\begin{figure}[h!]
  \centering
  \includegraphics[width=16.8cm]{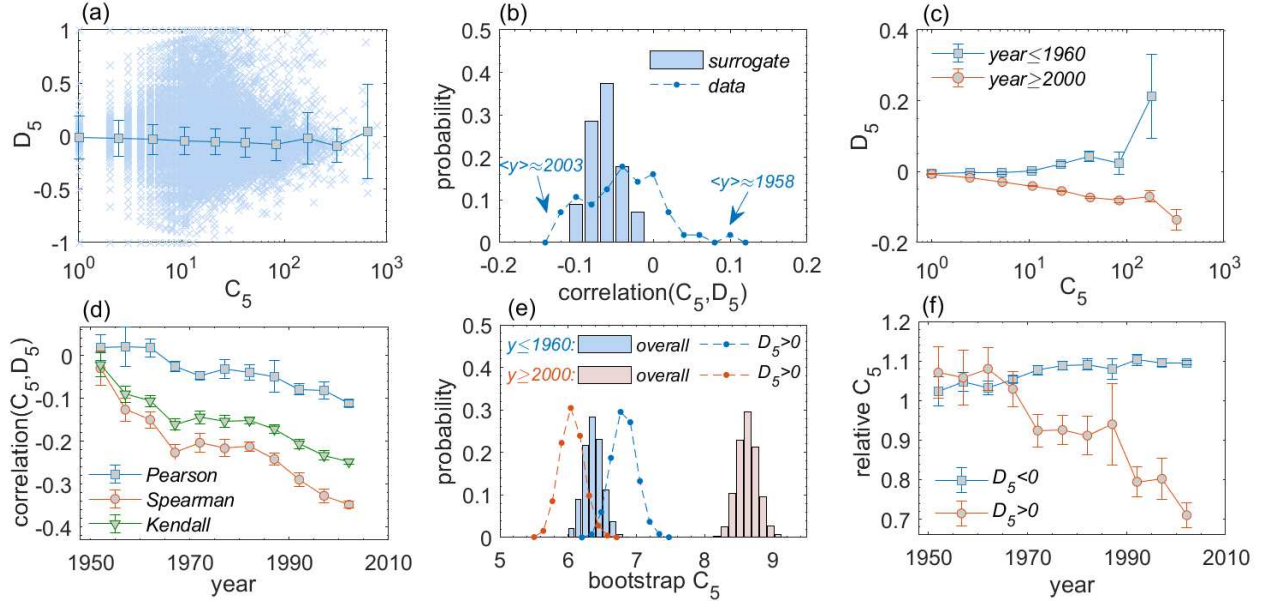}\\
  \caption{\textbf{Evolution of the correlation between disruption and citation.} (a) The scatter plot of the citation $C_5$ versus disruption $D_5$ values of papers, where each dot denotes the result of a single paper. The curve marks the average disruption $D_5$ of papers with different citations. (b) The distribution of the yearly (Pearson) correlation between citation $C_5$ and disruption $D_5$, where each correlation is computed by considering only papers published in a certain year. The arrows show the average year of the data points in the tails (i.e. $\leq-0.1$ or $\geq0.1$), respectively. The surrogate distribution which is much narrower is the result of the randomized case where the publication years of papers are randomly reshuffled. The Kolmogorov-Smirnov test of the distribution difference between the real and surrogate results in $p<0.001$. (c) The average disruption $D_5$ of papers having different citations, for papers published before 1960 and after 2000, respectively. (d) The evolution of the (Pearson, Spearman, Kendall) correlation between citation $C_5$ and disruption $D_5$, for papers published in different years. (e) The distributions of 10,000 realizations of bootstrap citation of papers published before 1960 (overall) and a subset of these papers with positive disruption ($D_5>0$), respectively. For comparison, we show also the distributions of 10,000 realizations of bootstrap citation of papers published after 2000 (overall) and a subset of these papers with positive disruption ($D_5>0$), respectively. The Kolmogorov-Smirnov tests of the distribution difference between the overall and $D_5>0$ result in $p<0.001$ for both $y\leq1960$ and $y\geq2000$. (f) The evolution of relative citations $C_5$ of $D_5<0$ and $D_5>0$ papers published in a year with respect to the mean citations of all the papers published in this year. Note the sharp decrease of citations for disruptive papers ($D_5>0$).}\label{fig2}
\end{figure}

\begin{figure}[h!]
  \centering
  \includegraphics[width=14cm]{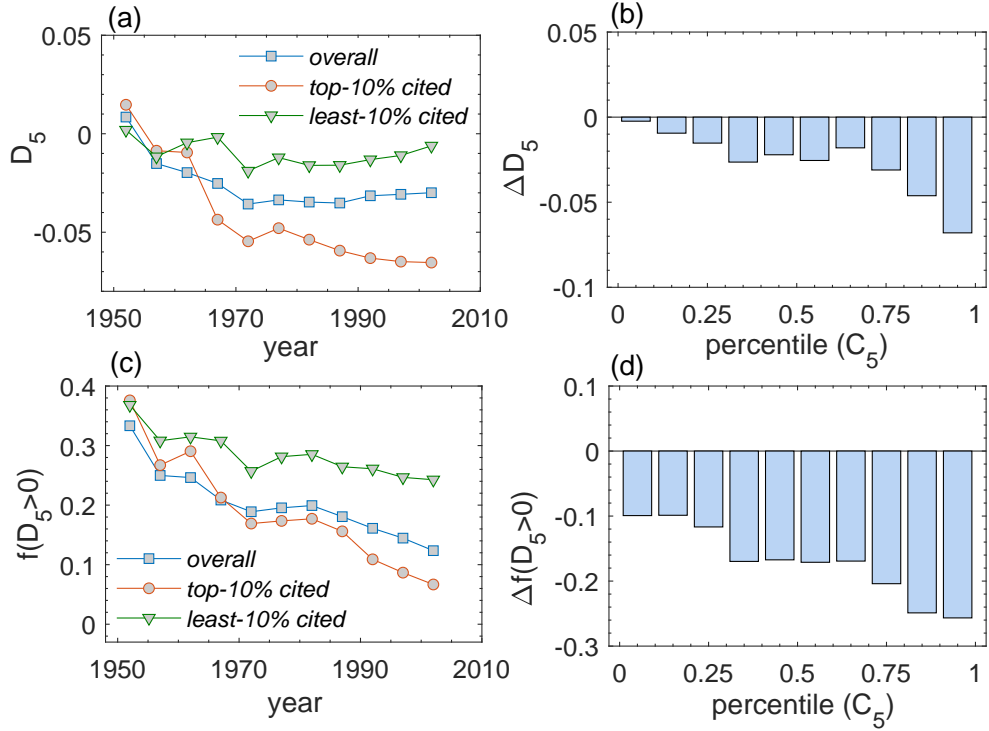}\\
  \caption{\textbf{Decreasing effect of disruption for papers with different impact.} (a) The evolution of the disruption of all the papers, 10\% most cited $C_5$ papers, and 10\% least cited $C_5$ papers in each year, respectively. (b) The $\Delta D_5$ (i.e. the difference between $\langle D_5\rangle$ after 2000 and $\langle D_5\rangle$ before 1960) for papers with different percentiles of $C_5$. (c) The fraction of papers with $D_5>0$ ,i.e. $f(D_5>0)$, for all the papers, 10\% most cited $C_5$ papers, and 10\% least cited $C_5$ papers in each year, respectively. (d) The $\Delta f(D_5>0)$ (i.e. the difference between $\langle f(D_5>0)\rangle$ after 2000 and $\langle f(D_5>0)\rangle$ before 1960) for papers with different percentiles of $C_5$.}\label{fig3}
\end{figure}

\begin{figure}[h!]
  \centering
  \includegraphics[width=16.8cm]{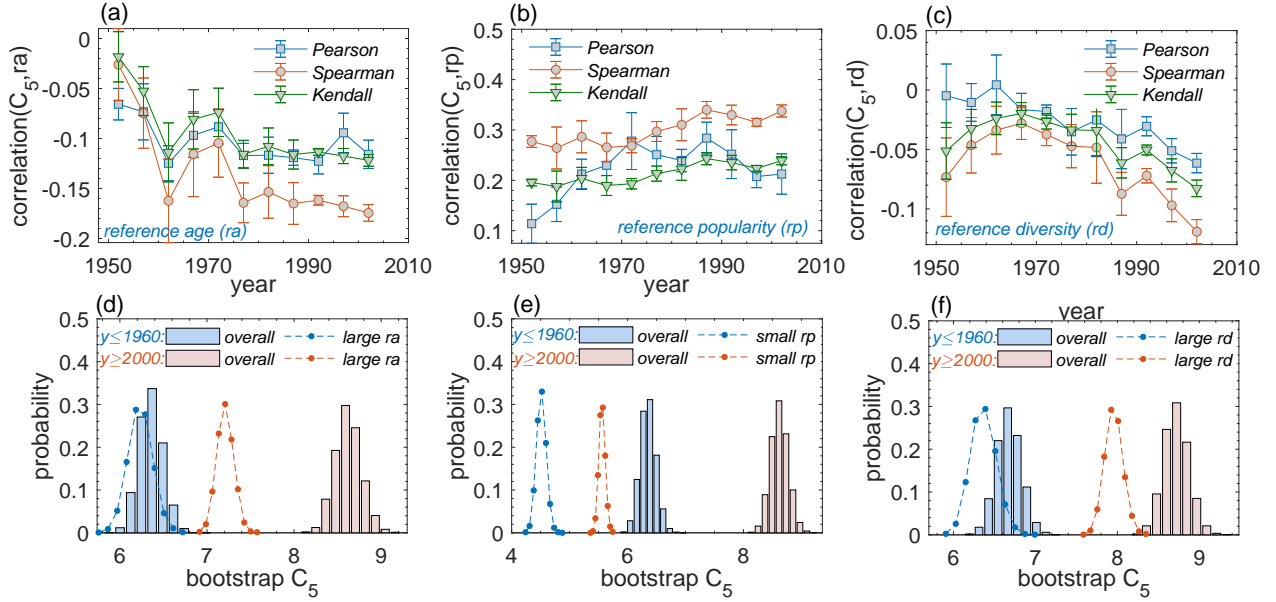}\\
  \caption{\textbf{Reference age, reference popularity and reference diversity as alternative measures of paper originality.} (a) Evolution of the correlation between reference age $ra$ and citation $C_5$, for papers published in different years. (b) Evolution of the correlation between reference popularity $rp$ and citation $C_5$, for papers published in different years. (c) Evolution of the correlation between reference diversity $rd$ and citation $C_5$, for papers published in different years. (d) The distributions of 10,000 realizations of bootstrap citation $C_5$ of papers published before 1960 (overall) and half of these papers with larger reference age, respectively. For comparison, we show also the distributions of 10,000 realizations of bootstrap citation $C_5$ of papers published after 2000 (overall) and half of these papers with largest reference age, respectively. The Kolmogorov-Smirnov tests of the distribution difference between the overall and large $ra$ result in $p<0.001$ for both $y\leq1960$ and $y\geq2000$. (e) and (f) are the same as (d), but for the results of low reference popularity and high reference diversity, respectively.}\label{fig4}
\end{figure}

\begin{figure}[h!]
  \centering
  \includegraphics[width=16.8cm]{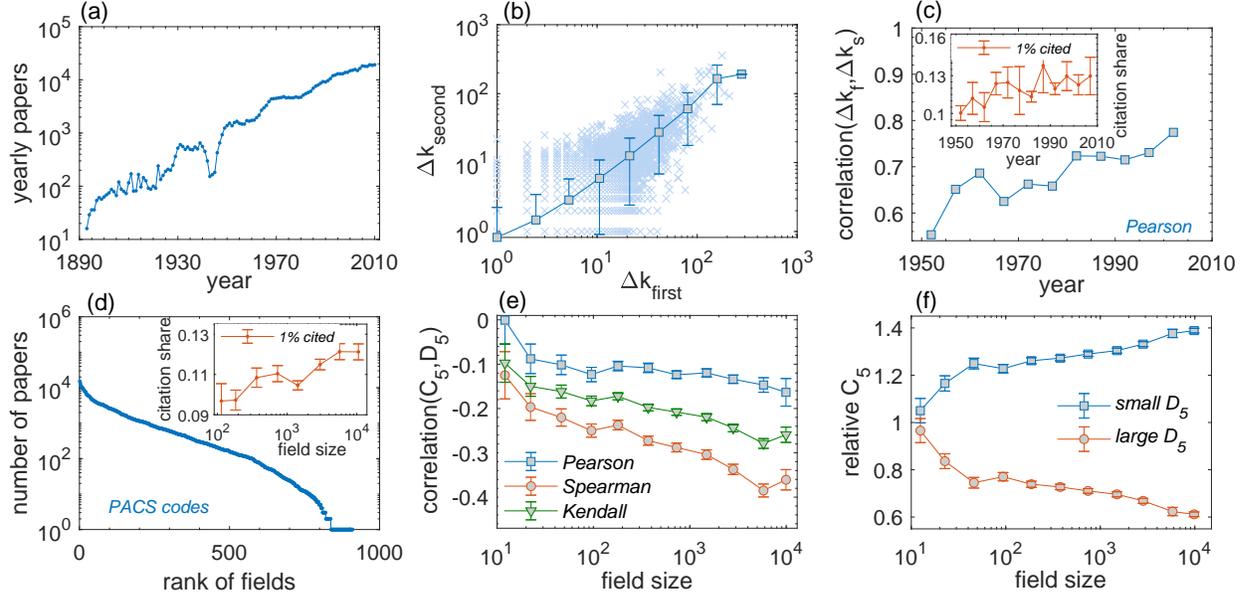}\\
  \caption{\textbf{Information overload and decreasing attention to disruptive works in large fields.} (a) The exponentially increasing number of published papers during the years. (b) We take the papers published in 2000, and make the scatter plot of the citations they received in the first 5 years after publication $\Delta k_{\rm first}$ versus the citations received in the second 5 years after publication $\Delta k_{\rm second}$. The curve shows the average. The positive correlation suggests a preferential attachment. (c) The evolution of correlation between $\Delta k_{\rm first}$ and $\Delta k_{\rm second}$ for papers published in different years. The increasing correlation coefficient suggests increasingly stronger effect of preferential attachment. The inset shows the evolution of the fraction of citations of top-1\% $C_5$ most highly cited papers published in a year. (d) The Zipf plot of the number of papers for each PACS code (a research field). Inset is the fraction of citations of top-1\% $C_5$ most highly cited papers in fields of different sizes. (e) The correlations between citation $C_5$ and disruption $D_5$, for papers published in fields of different sizes. (f) The relative citations $C_5$ of the 50\% papers with larger $D_5$ and the 50\% papers with smaller $D_5$ in fields of different sizes.}\label{fig5}
\end{figure}

\begin{figure}[h!]
  \centering
  \includegraphics[width=16cm]{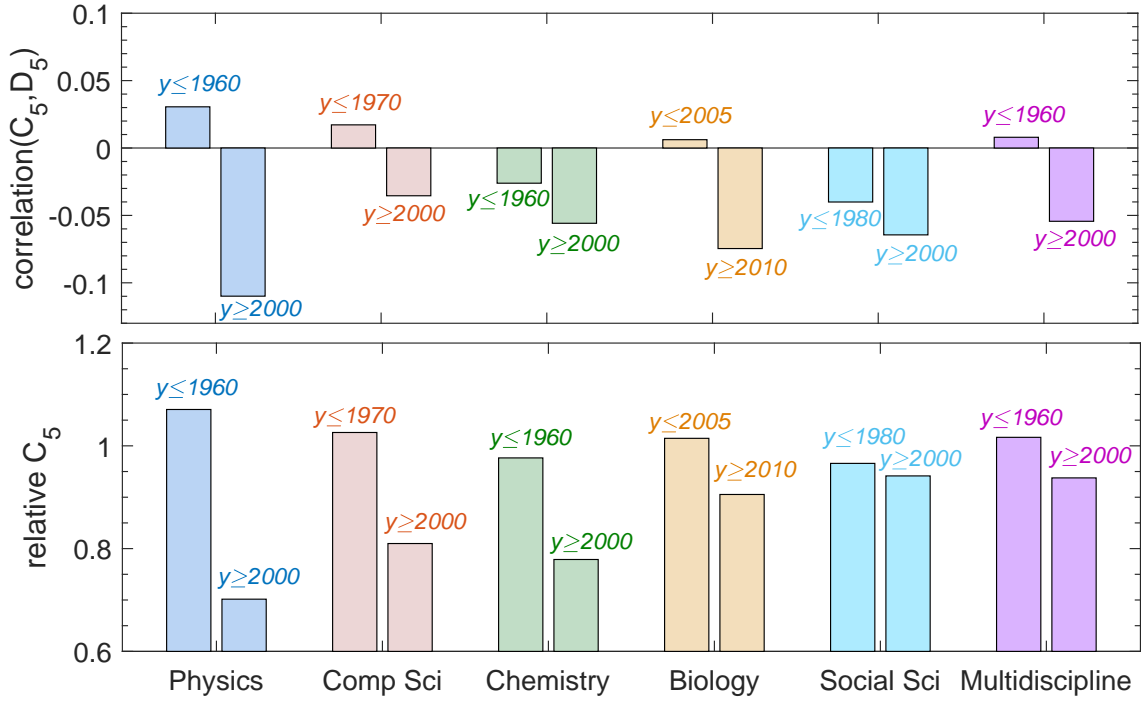}\\
  \caption{\textbf{Discipline comparison.} Upper panel: the (Pearson) correlation between citation $C_5$ and disruption $D_5$, for the papers published in different fields in early years and in recent years, respectively. Bottom panel: the relative citations $C_5$ of the papers with positive disruption $D_5>0$ with respect to the mean citations of all papers, for the papers published in different fields in early years and in recent years, respectively.}\label{fig6}
\end{figure}

\clearpage
\begin{center}
{\large\bfseries Supplementary Information}\\[8pt]
{\large Disruptive works in science are losing impact}\\[8pt]
\small An Zeng, Ying Fan, Zengru Di, Yougui Wang and Shlomo Havlin\\
\end{center}


\begin{figure}[h!]
  \centering
  \includegraphics[width=16.5cm]{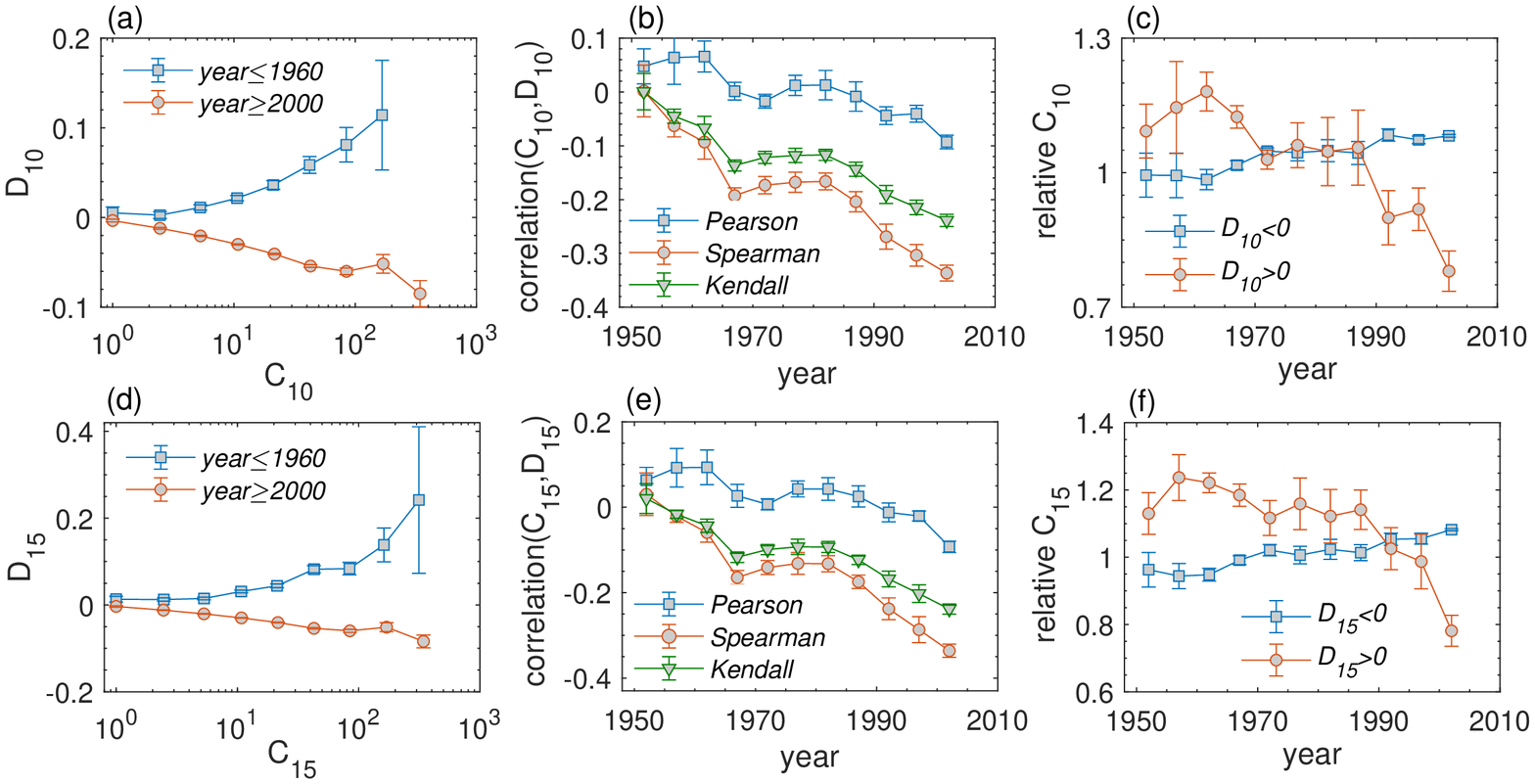}\\
  \textbf{Supplementary Figure 1. The results of $D_{10}$ and $D_{15}$.} (a) The average disruption $D_{10}$ of papers having different citations $C_{10}$, for papers published before 1960 and after 2000, respectively. (b) The evolution of the (Pearson, Spearman, Kendall) correlation between citation $C_{10}$ and disruption $D_{10}$, for papers published in different years. (c) The evolution of relative citations $C_{10}$ of $D_{10}<0$ and $D_{10}>0$ papers published in a year with respect to the mean citations of all the papers published in this year. (d) The average disruption $D_{15}$ of papers having different citations $C_{15}$, for papers published before 1960 and after 2000, respectively. (e) The evolution of the (Pearson, Spearman, Kendall) correlation between citation $C_{15}$ and disruption $D_{15}$, for papers published in different years. (f) The evolution of relative citations $C_{15}$ of $D_{15}<0$ and $D_{15}>0$ papers published in a year with respect to the mean citations of all the papers published in this year. Both of the results of $D_{10}$ and $D_{15}$ show similar trends as those of $D_5$ presented in the main paper. \label{FigS}
\end{figure}

\clearpage
\begin{figure}[h!]
  \centering
  \includegraphics[width=16.5cm]{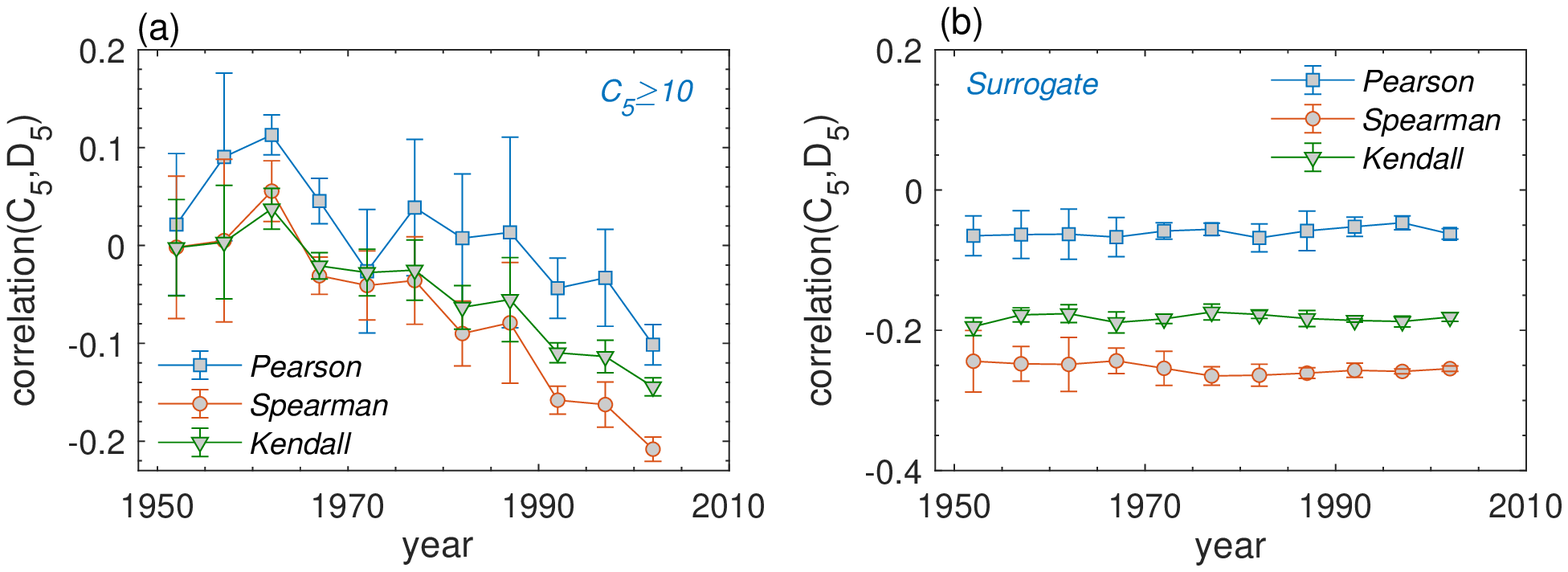}\\
  \textbf{Supplementary Figure 2. Avoiding extreme values of disruption by excluding papers with very few citations.} (a) Papers with few citations may have extreme values of disruption (i.e. 1 or -1). To avoid this effect on the general trend, we calculate here the evolution of the correlation between disruption $D_5$ and citation $C_5$ only for papers with at least 10 citations. (b) The evolution of the correlation between citation $C_5$ and disruption $D_5$ in a control surrogate where the publication years of papers are randomly reshuffled. The flat curves here suggest that the decreasing correlation between citation and disruption cannot be explained by random behaviours. \label{FigS}
\end{figure}

\clearpage
\begin{figure}[h!]
  \centering
  \includegraphics[width=14cm]{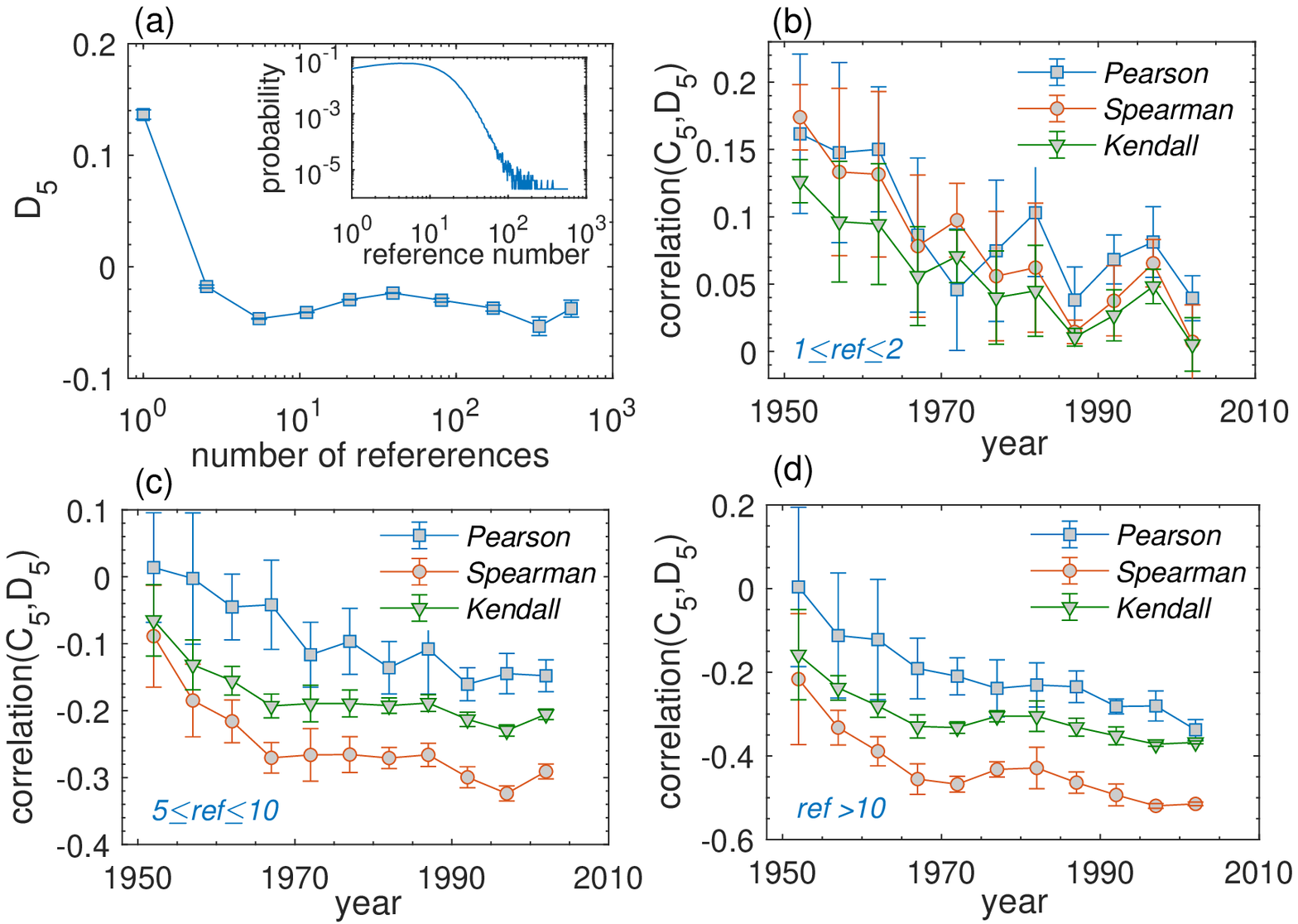}\\
  \textbf{Supplementary Figure 3. Controlling the number of references of focal papers.} (a) The mean disruption $D_5$ of papers with different number of references. The result suggests that the disruption value strongly depends on the number of references that a paper has (i.e. fewer references, higher disruption). The inset shows the distribution of the number of references of papers. To remove the effect of references on disruption, here we redo the analysis in the paper with the number of papers' references controlled. (b) The evolution of the (Pearson, Spearman, Kendall) correlation between citation $C_{5}$ and disruption ($D_{5}$), for papers with 1 or 2 references. (c) The evolution of the (Pearson, Spearman, Kendall) correlation between citation $C_{5}$ and disruption ($D_{5}$), for papers with 5 to 10 references. (d) The evolution of the (Pearson, Spearman, Kendall) correlation between citation $C_{5}$ and disruption ($D_{5}$), for papers with more than 10 references. The trends here are consistent with those presented in the main paper.\label{FigS}
\end{figure}

\clearpage
\begin{figure}[h!]
  \centering
  \includegraphics[width=16.5cm]{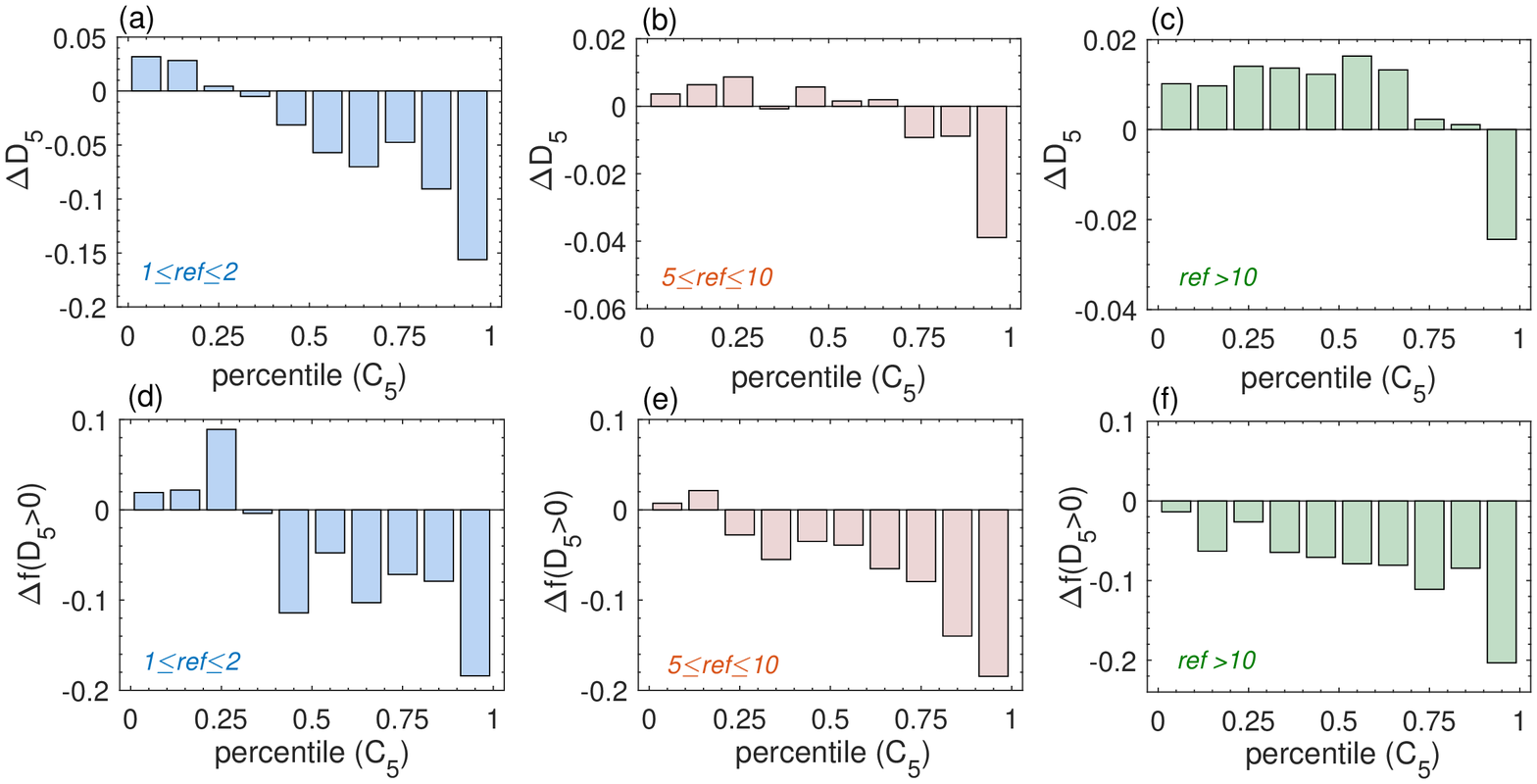}\\
  \textbf{Supplementary Figure 4. The decreasing disruption over time when the reference number of papers are controlled.} As the number of references of a paper is an important factor affecting its disruption value, we study here the evolution of the disruption of impactful and less impactful papers, taking into account only the papers with similar references in the APS data. (a-c) The $\Delta D_5$ (i.e. the difference between $\langle D_5\rangle$ after 2000 and $\langle D_5\rangle$ before 1960) for papers with different percentiles of $C_5$. Here, (a) only includes papers with 1 or 2 references. (b) only includes papers with 5 to 10 references. (c) only includes papers with more than 10 references. (d-f) The $\Delta f(D_5>0)$ (i.e. the difference between $\langle f(D_5>0)\rangle$ after 2000 and $\langle f(D_5>0)\rangle$ before 1960) for papers with different percentiles of $C_5$. (d) only includes papers with 1 or 2 references. (e) only includes papers with 5 to 10 references. (f) only includes papers with more than 10 references. The trends here are consistent with those presented in the main paper.\label{FigS}
\end{figure}

\clearpage
\begin{figure}[h!]
  \centering
  \includegraphics[width=16.5cm]{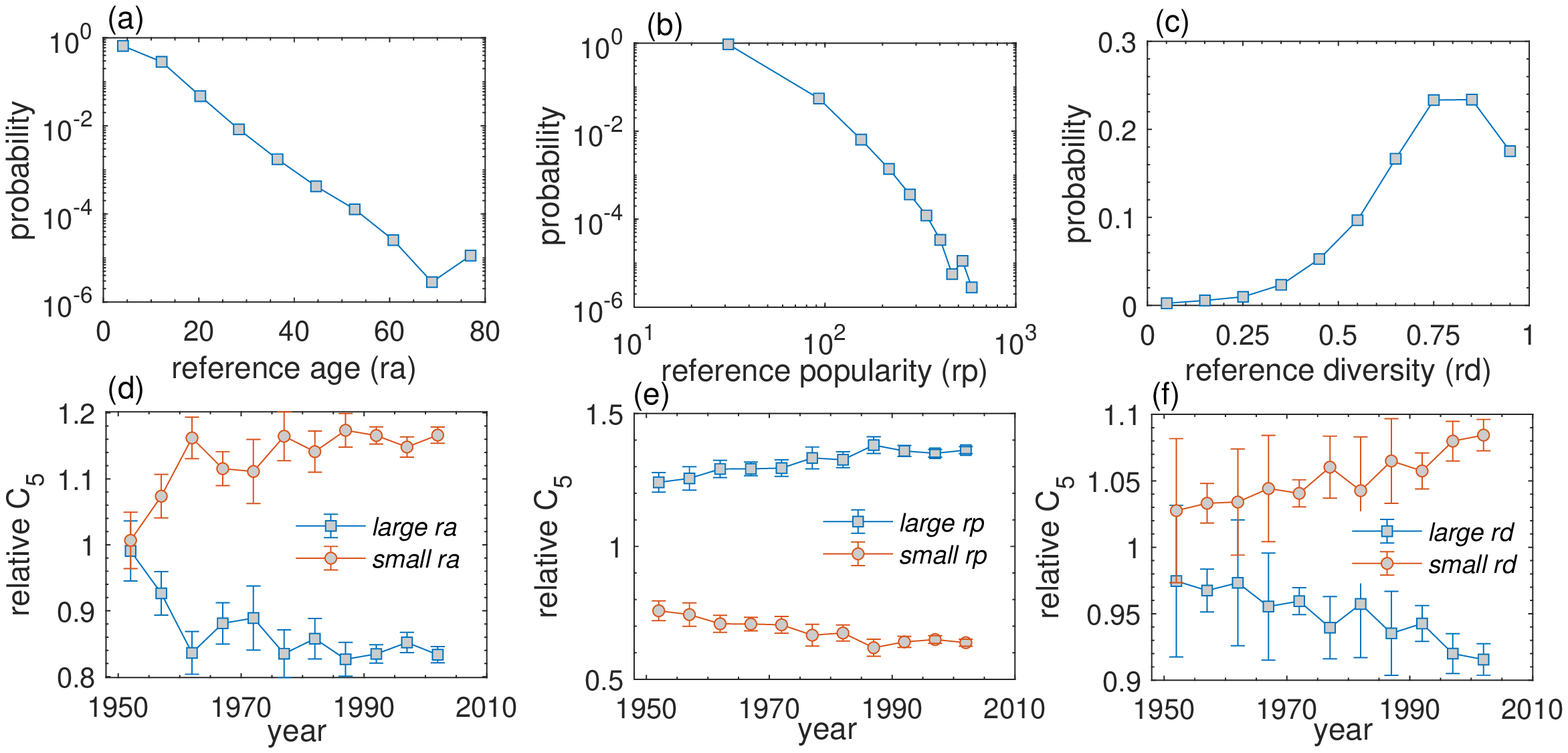}\\
  \textbf{Supplementary Figure 5. Metrics based on references.} In the main paper, we consider three other metrics that have been shown to connect to paper originality. Unlike the disruption metric relying on both the information of references and citing papers, their calculation solely depends on paper references (thus completely independent of citations). These metrics are reference age, reference popularity and reference diversity. (a) The distribution of the reference age ($ra$) of papers. (b) The distribution of the reference popularity $rp$ of papers. (c) The distribution of the reference diversity $rd$ of papers (see the main text for the definition of these metrics). (d) The relative citations $C_5$ of the 50\% papers with larger $ra$ and the 50\% papers with smaller $ra$ in a year with respect to the mean citations of all the papers published in this year. (e) The relative citations $C_5$ of the 50\% papers with larger $rp$ and the 50\% papers with smaller $rp$ in a year with respect to the mean citations of all the papers published in this year. (f) The relative citations $C_5$ of the 50\% papers with larger $rd$ and the 50\% papers with smaller $rd$ in a year with respect to the mean citations of all the papers published in this year. The results suggest that papers with large $ra$, small $rp$ and large $rd$ keep losing citations over time.\label{FigS}
\end{figure}

\clearpage
\begin{figure}[h!]
  \centering
  \includegraphics[width=14cm]{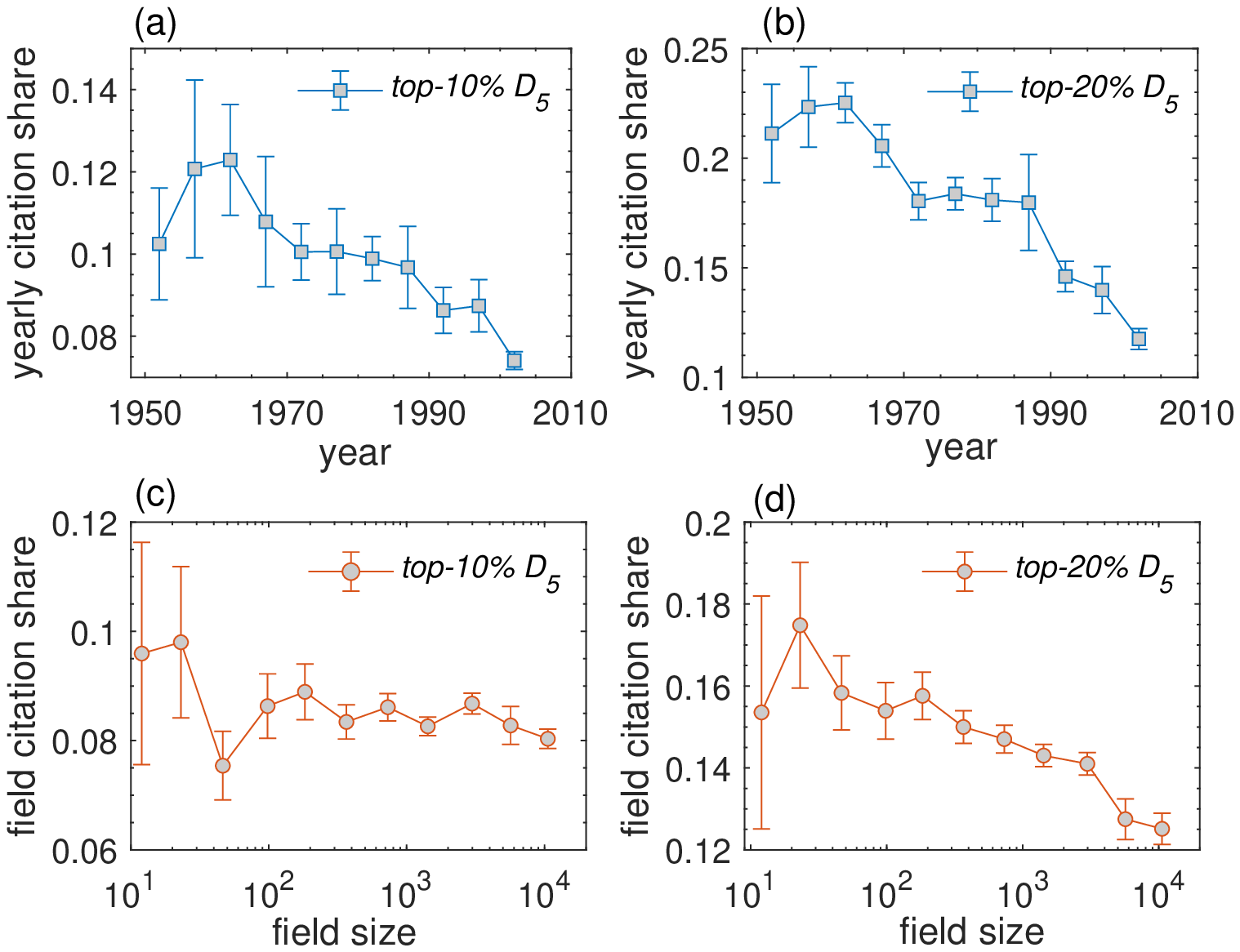}\\
  \textbf{Supplementary Figure 6. Citation share of highly disruptive papers in each year.} (a) The evolution of the citation $C_5$ share of top-10\% most highly disruptive papers in $D_5$ published in a year. (b) The evolution of the citation $C_5$ share of top-20\% most highly disruptive papers in $D_5$ published in a year. (c) The evolution of the citation $C_5$ share of top-10\% most highly disruptive papers in $D_5$ published in fields of different sizes. (d) The evolution of the citation $C_5$ share of top-20\% most highly disruptive papers in $D_5$ published in fields of different sizes. One can see a clear decreasing trend in all cases, suggesting that the highly disruptive papers have fewer and fewer citation share over time or as fields get larger.\label{FigS}
\end{figure}

\clearpage
\begin{figure}[h!]
  \centering
  \includegraphics[width=16.5cm]{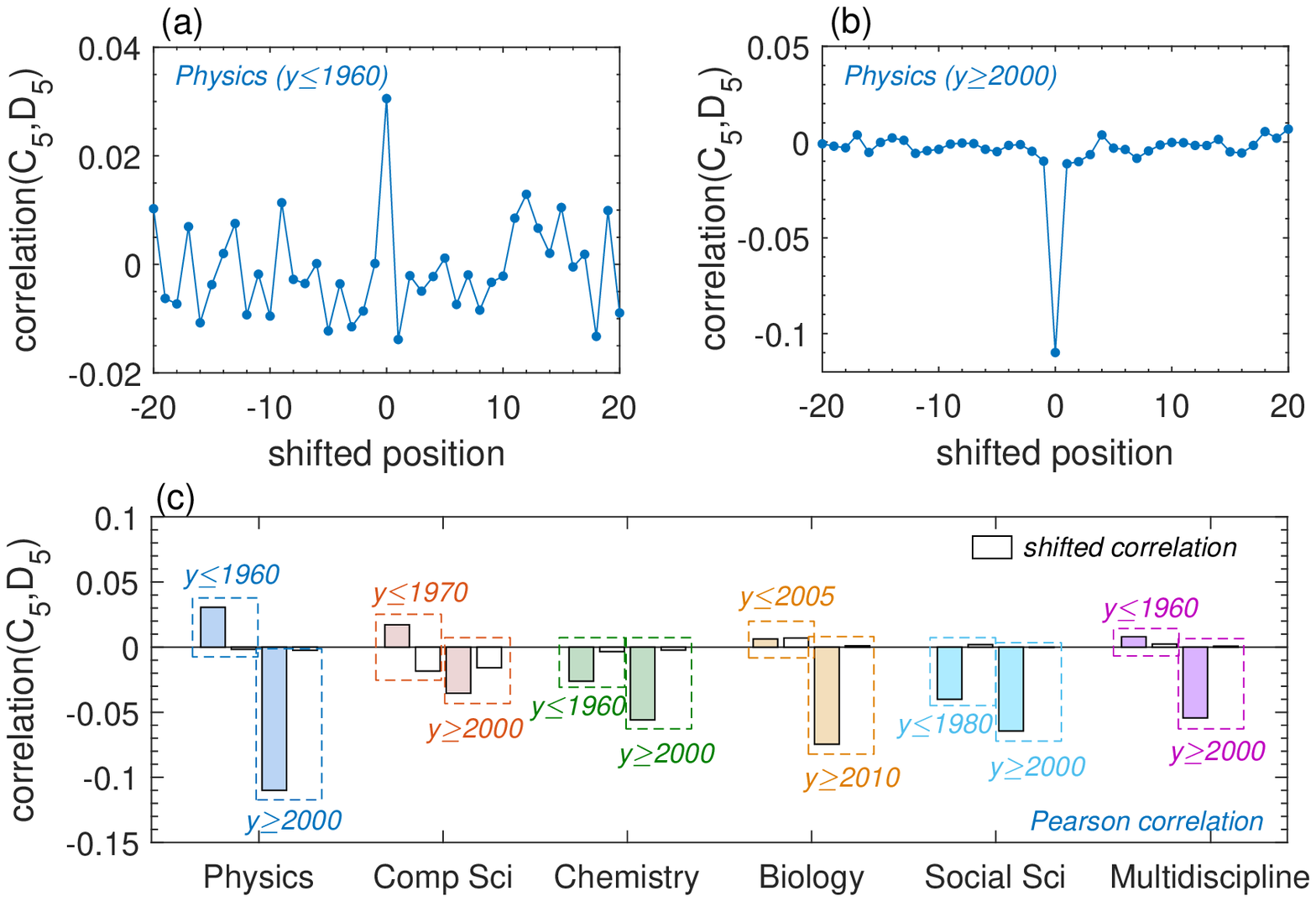}\\
  \textbf{Supplementary Figure 7. Significance of the correlations between citation and disruption.} To support the significance of the correlation between citation $C_5$ and disruption $D_5$ in Fig. 6 in the paper, we conduct the Pearson shifting test in which the Pearson correlation is calculated after shifting all elements in the vector of disruption $D_5$ by certain number of positions. (a) We take all papers in APS before 1960 and calculate the Pearson correlation (between the vectors of citation and disruption) with shifting one vector with respect to the other by 20 positions (-20 to 20). The true correlation is without movements while each movement is like a random correlations, showing therefore the level of noise. The sharp peak at shifting zero suggests that the correlation of citation and disruption in the original data is indeed significant. (b) The same test for papers in APS after 2000. Similar sharp peak at shifting zero can be observed. (c) The Pearson shifting test for data from different disciplines. The colored bars are the results of real data, while the bars with no color are the results of the average correlation with shifted positions from -20 to 20, excluding the case of zero shifted position (i.e. the real data). The nearly zero values of the shifted Pearson correlations suggest that the observed correlations in real data are indeed significant.\label{FigS}
\end{figure}

\clearpage
\begin{figure}[h!]
  \centering
  \includegraphics[width=16.5cm]{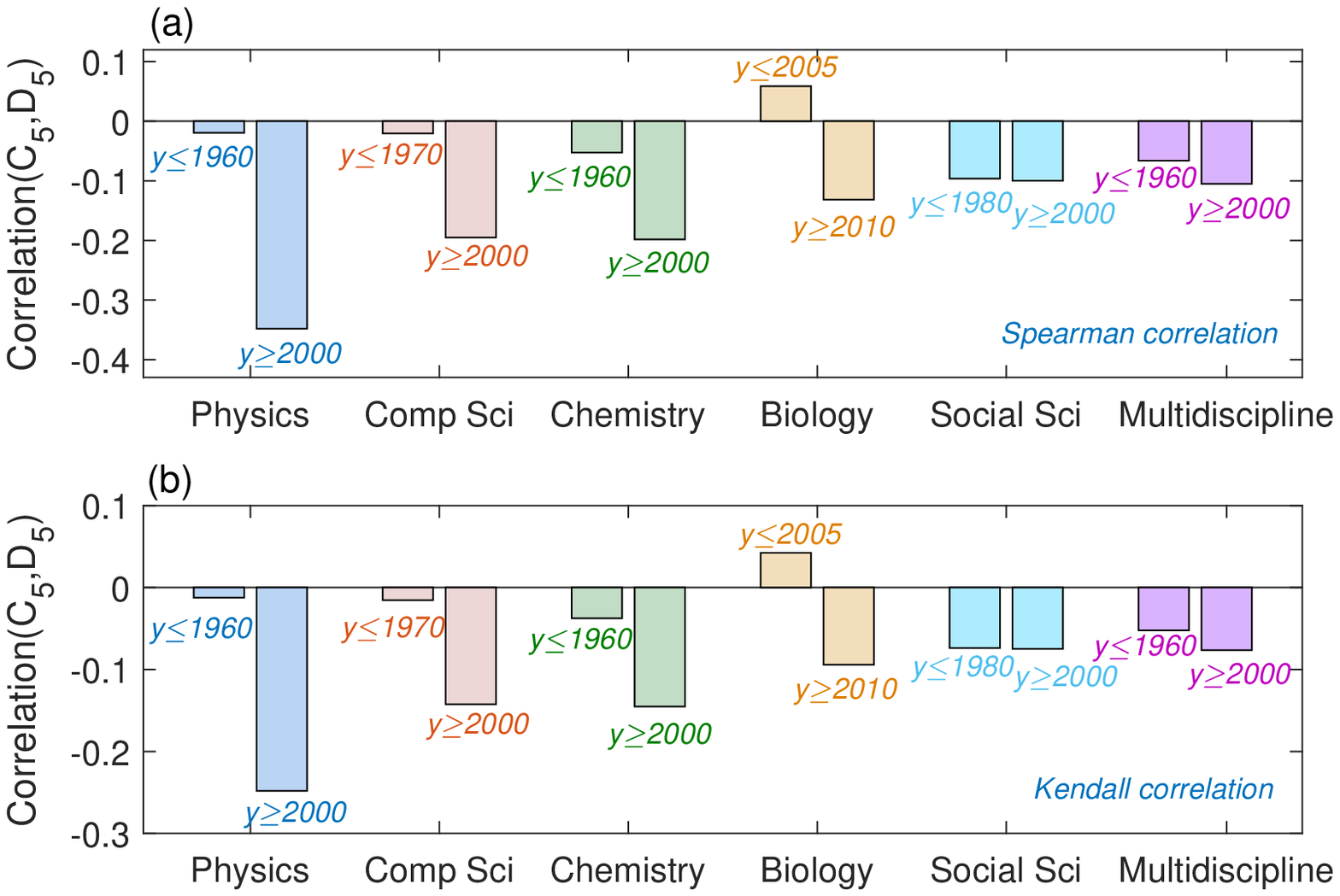}\\
  \textbf{Supplementary Figure 8. Spearman and Kendall correlation between citation and disruption in data from different disciplines.} (a) The Spearman rank correlation between citation $C_5$ and disruption $D_5$, for the papers published in different fields in early years and in recent years, respectively. (b) The Kendall rank correlation between citation $C_5$ and disruption $D_5$, for the papers published in different fields in early years and in recent years, respectively. In both panels, one can observe a decreasing trend of the correlations in all disciplines. \label{FigS}
\end{figure}

\clearpage
\begin{figure}[h!]
  \centering
  \includegraphics[width=16.5cm]{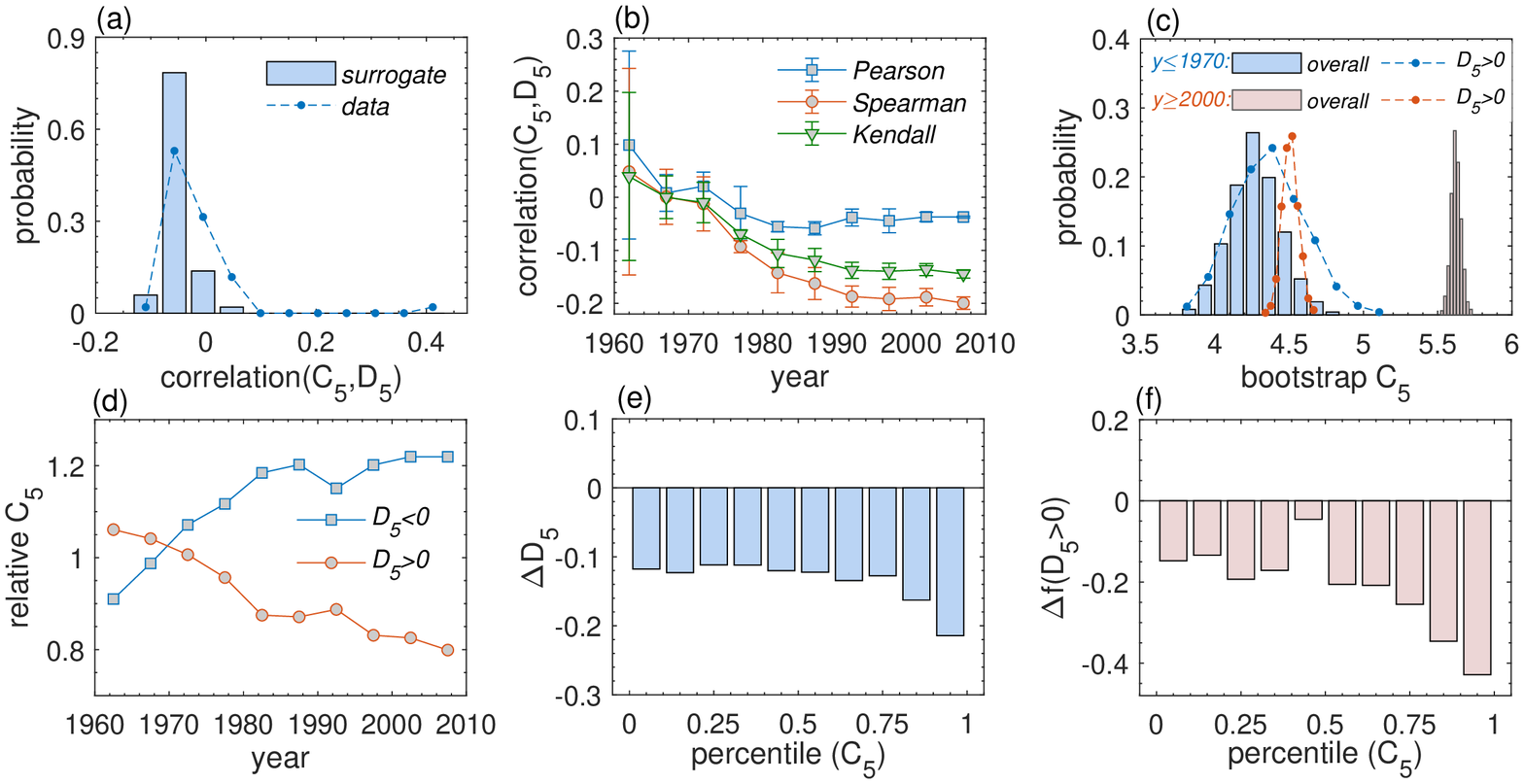}\\
  \textbf{Supplementary Figure 9. Results of computer science data.} (a) The distribution of the yearly (Pearson) correlation between citation $C_5$ and disruption $D_5$, with each correlation computed by only taking papers published in a certain year. The surrogate distribution which is much narrower is the result of the randomized case where the publication years of papers are randomly reshuffled (b) The evolution of the (Pearson, Spearman, Kendall) correlation between disruption $D_5$ and citation $C_5$, for papers published in different years. (c) The distributions of 10,000 realizations of bootstrap citation $C_5$ of papers published before 1970 (overall) and a subset of these papers with positive disruption ($D_{5}>0$), respectively. For comparison, we show also the distributions of 10,000 realizations of bootstrap citation of papers published after 2000 (overall) and a subset of these papers with positive disruption ($D_{5}>0$), respectively. (d) The evolution of relative citations $C_5$ of $D_{5}<0$ and $D_{5}>0$ papers published in a year with respect to the mean citations $C_5$ of all the papers published in this year. (e) The $\Delta D_5$ (i.e. the difference between $\langle D_5\rangle$ after 2000 and $\langle D_5\rangle$ before 1970) for papers with different percentiles of $C_5$. (f) The $\Delta f(D_5>0)$ (i.e. the difference between $\langle f(D_5>0)\rangle$ after 2000 and $\langle f(D_5>0)\rangle$ before 1970) for papers with different percentiles of $C_5$. \label{FigS}
\end{figure}

\clearpage
\begin{figure}[h!]
  \centering
  \includegraphics[width=16.5cm]{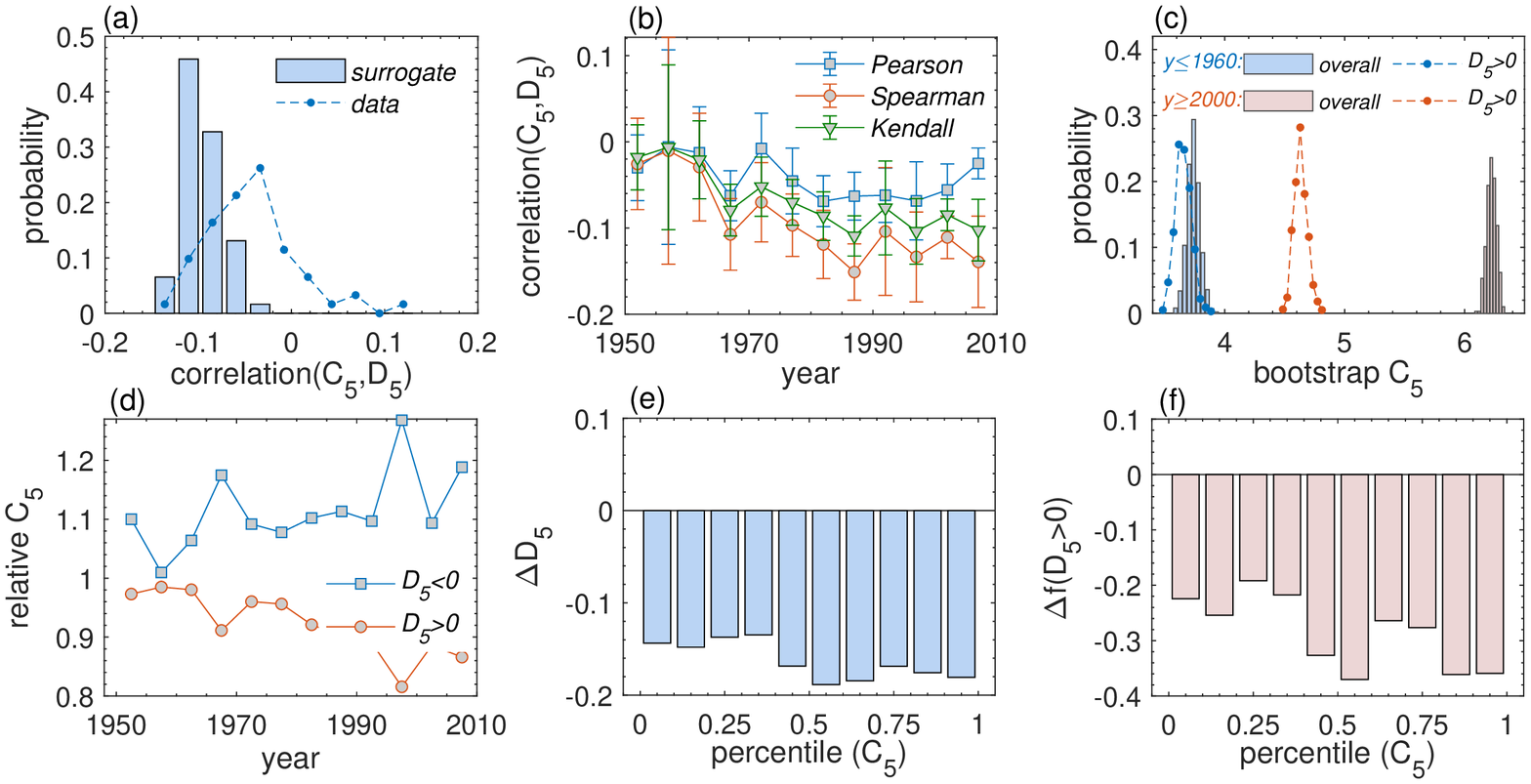}\\
  \textbf{Supplementary Figure 10. Results of Chemistry data.} (a) The distribution of the yearly (Pearson) correlation between citation $C_5$ and disruption $D_5$, with each correlation computed by only taking papers published in a certain year. The surrogate distribution which is much narrower is the result of the randomized case where the publication years of papers are randomly reshuffled (b) The evolution of the (Pearson, Spearman, Kendall) correlation between disruption $D_5$ and citation $C_5$, for papers published in different years. (c) The distributions of 10,000 realizations of bootstrap citation $C_5$ of papers published before 1960 (overall) and a subset of these papers with positive disruption ($D_{5}>0$), respectively. For comparison, we show also the distributions of 10,000 realizations of bootstrap citation of papers published after 2000 (overall) and a subset of these papers with positive disruption ($D_{5}>0$), respectively. (d) The evolution of relative citations $C_5$ of $D_{5}<0$ and $D_{5}>0$ papers published in a year with respect to the mean citations $C_5$ of all the papers published in this year. (e) The $\Delta D_5$ (i.e. the difference between $\langle D_5\rangle$ after 2000 and $\langle D_5\rangle$ before 1960) for papers with different percentiles of $C_5$. (f) The $\Delta f(D_5>0)$ (i.e. the difference between $\langle f(D_5>0)\rangle$ after 2000 and $\langle f(D_5>0)\rangle$ before 1960) for papers with different percentiles of $C_5$. \label{FigS}
\end{figure}

\clearpage
\begin{figure}[h!]
  \centering
  \includegraphics[width=16.5cm]{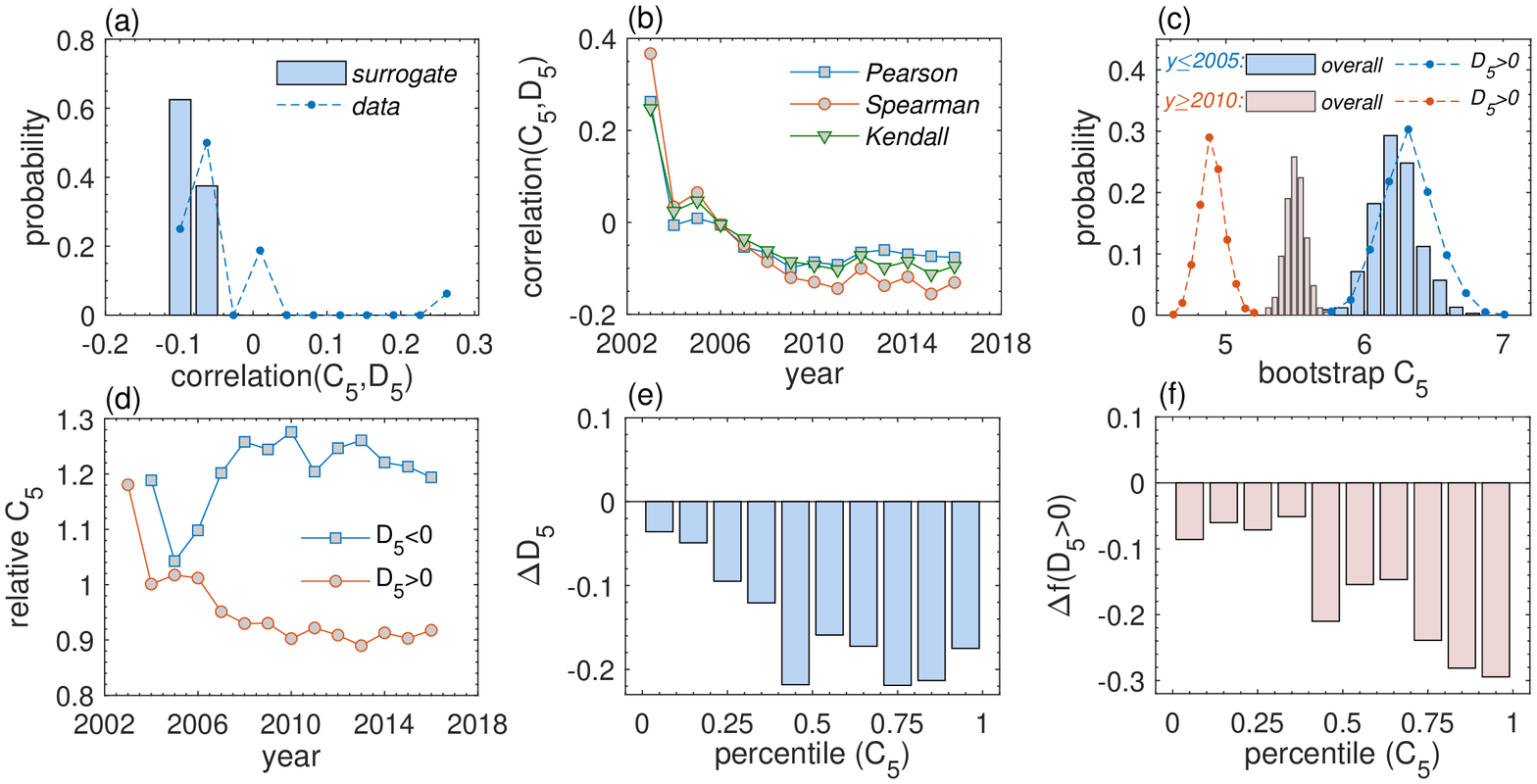}\\
  \textbf{Supplementary Figure 11. Results of Biology data.} (a) The distribution of the yearly (Pearson) correlation between citation $C_5$ and disruption $D_5$, with each correlation computed by only taking papers published in a certain year. The surrogate distribution which is much narrower is the result of the randomized case where the publication years of papers are randomly reshuffled (b) The evolution of the (Pearson, Spearman, Kendall) correlation between disruption $D_5$ and citation $C_5$, for papers published in different years. (c) The distributions of 10,000 realizations of bootstrap citation $C_5$ of papers published before 2005 (overall) and a subset of these papers with positive disruption ($D_{5}>0$), respectively. For comparison, we show also the distributions of 10,000 realizations of bootstrap citation of papers published after 2010 (overall) and a subset of these papers with positive disruption ($D_{5}>0$), respectively. (d) The evolution of relative citations $C_5$ of $D_{5}<0$ and $D_{5}>0$ papers published in a year with respect to the mean citations $C_5$ of all the papers published in this year. (e) The $\Delta D_5$ (i.e. the difference between $\langle D_5\rangle$ after 2010 and $\langle D_5\rangle$ before 2005) for papers with different percentiles of $C_5$. (f) The $\Delta f(D_5>0)$ (i.e. the difference between $\langle f(D_5>0)\rangle$ after 2010 and $\langle f(D_5>0)\rangle$ before 2005) for papers with different percentiles of $C_5$. \label{FigS}
\end{figure}

\clearpage
\begin{figure}[h!]
  \centering
  \includegraphics[width=16.5cm]{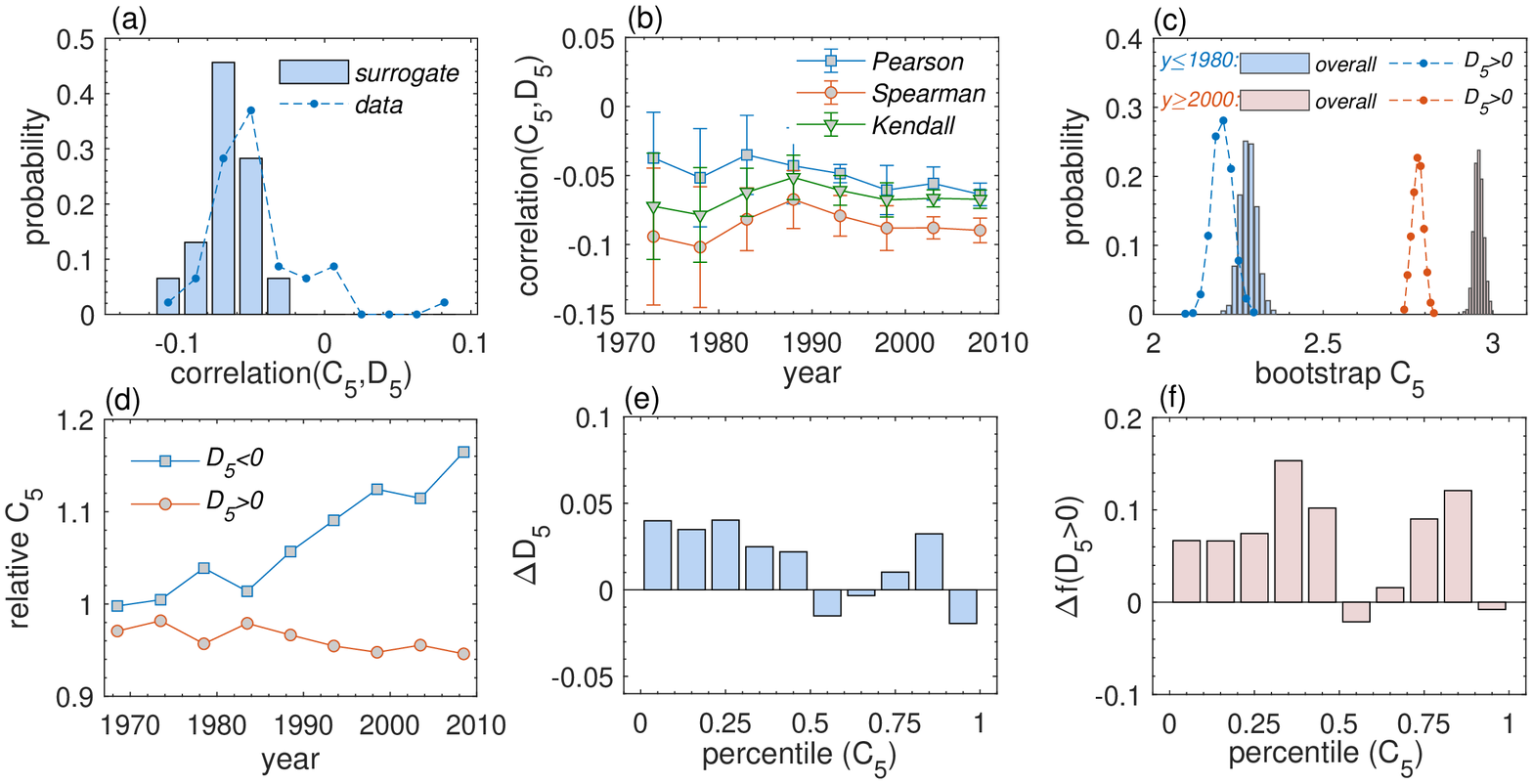}\\
  \textbf{Supplementary Figure 12. Results of Social science data.} (a) The distribution of the yearly (Pearson) correlation between citation $C_5$ and disruption $D_5$, with each correlation computed by only taking papers published in a certain year. The surrogate distribution which is much narrower is the result of the randomized case where the publication years of papers are randomly reshuffled (b) The evolution of the (Pearson, Spearman, Kendall) correlation between disruption $D_5$ and citation $C_5$, for papers published in different years. (c) The distributions of 10,000 realizations of bootstrap citation $C_5$ of papers published before 1980 (overall) and a subset of these papers with positive disruption ($D_{5}>0$), respectively. For comparison, we show also the distributions of 10,000 realizations of bootstrap citation of papers published after 2000 (overall) and a subset of these papers with positive disruption ($D_{5}>0$), respectively. (d) The evolution of relative citations $C_5$ of $D_{5}<0$ and $D_{5}>0$ papers published in a year with respect to the mean citations $C_5$ of all the papers published in this year. (e) The $\Delta D_5$ (i.e. the difference between $\langle D_5\rangle$ after 2000 and $\langle D_5\rangle$ before 1980) for papers with different percentiles of $C_5$. (f) The $\Delta f(D_5>0)$ (i.e. the difference between $\langle f(D_5>0)\rangle$ after 2000 and $\langle f(D_5>0)\rangle$ before 1980) for papers with different percentiles of $C_5$. \label{FigS}
\end{figure}

\clearpage
\begin{figure}[h!]
  \centering
  \includegraphics[width=16.5cm]{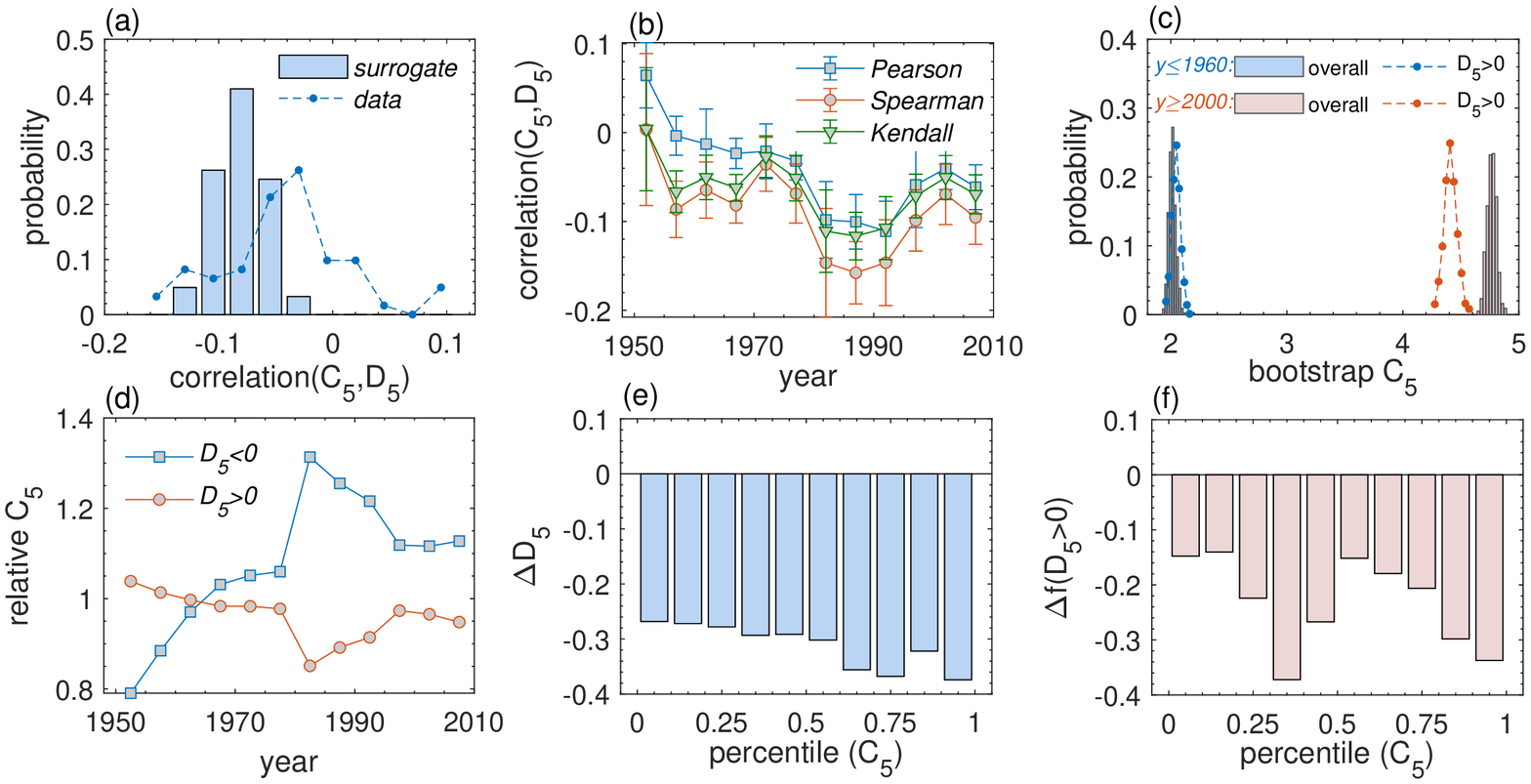}\\
  \textbf{Supplementary Figure 13. Results of Multidisciplinary science data.} ((a) The distribution of the yearly (Pearson) correlation between citation $C_5$ and disruption $D_5$, with each correlation computed by only taking papers published in a certain year. The surrogate distribution which is much narrower is the result of the randomized case where the publication years of papers are randomly reshuffled (b) The evolution of the (Pearson, Spearman, Kendall) correlation between disruption $D_5$ and citation $C_5$, for papers published in different years. (c) The distributions of 10,000 realizations of bootstrap citation $C_5$ of papers published before 1960 (overall) and a subset of these papers with positive disruption ($D_{5}>0$), respectively. For comparison, we show also the distributions of 10,000 realizations of bootstrap citation of papers published after 2000 (overall) and a subset of these papers with positive disruption ($D_{5}>0$), respectively. (d) The evolution of relative citations $C_5$ of $D_{5}<0$ and $D_{5}>0$ papers published in a year with respect to the mean citations $C_5$ of all the papers published in this year. (e) The $\Delta D_5$ (i.e. the difference between $\langle D_5\rangle$ after 2000 and $\langle D_5\rangle$ before 1960) for papers with different percentiles of $C_5$. (f) The $\Delta f(D_5>0)$ (i.e. the difference between $\langle f(D_5>0)\rangle$ after 2000 and $\langle f(D_5>0)\rangle$ before 1960) for papers with different percentiles of $C_5$.\label{FigS}
\end{figure}

\end{document}